\documentclass[letterpaper,twocolumn,aps,prb,groupedaddress,floatfix]{revtex4-2}

\usepackage[USenglish]{babel}
\usepackage[utf8]{inputenc}
\usepackage[dvipsnames, hyperref]{xcolor}
\usepackage[normalem]{ulem}

\usepackage{multirow}
\usepackage{mathtools}
\usepackage{amssymb}
\usepackage{amsbsy}
\usepackage{amsmath}
\usepackage{amsfonts}
\usepackage{stmaryrd}
\usepackage{array}
\usepackage{bm}
\usepackage[arrowdel]{physics}
\usepackage{units}

\usepackage{graphicx}
\usepackage{epsfig}
\usepackage[above, below]{placeins}
\usepackage{tikz}
\usetikzlibrary{quantikz}  
\usetikzlibrary{braids}

\usepackage{natbib}
\usepackage{hyperref}
\hypersetup{colorlinks,linkcolor=ForestGreen,citecolor=ForestGreen,urlcolor=ForestGreen, pdfauthor={Jared Jeyaretnam, Chris J.\ Turner, Arijeet Pal},pdftitle={Renormalization view on resonance proliferation between many-body localized phases}}

\def\rsrgx#{RSRG\babelhyphen{nobreak}X}  
\newcommand*{\s}[1]{\sigma^{\MakeLowercase{#1}}}

\makeatletter

\begin{document}
\title{Renormalization view on resonance proliferation between many-body localized phases}

\author{Jared Jeyaretnam}
\affiliation{Department of Physics and Astronomy, University College London,
	Gower Street, London WC1E 6BT, UK}

\author{Christopher J. Turner}
\affiliation{Department of Physics and Astronomy, University College London,
	Gower Street, London WC1E 6BT, UK}

\author{Arijeet Pal}
\affiliation{Department of Physics and Astronomy, University College London,
	Gower Street, London WC1E 6BT, UK}

\begin{abstract}
	\noindent
 	Topology and many-body localization (MBL) have opened new avenues for preserving quantum information at finite energy density.
 	Resonant delocalization plays a crucial role in destabilizing these phenomena.
 	In this work, we study the statistical properties of many-body resonances in a disordered interacting Ising model -- which can host symmetry protected topological order -- using a Clifford circuit encoding of the real space renormalization group which allows the resonant properties of the wave functions to be efficiently characterized.
 	Our findings show that both the trivial and topologically ordered MBL phases remain stable to the resonances, but in the vicinity of the transition between them localization is destabilized by resonance proliferation.
 	Diverging susceptibility towards the development of an avalanche instability suggests an intervening ergodic phase.
 	We are also able to access the local integrals of motion in the MBL phases and identify the topological edge-mode operators in the ordered phase.
 	Our results have important implications for the stability of MBL and phase transitions between distinct MBL phases with and without symmetries.
\end{abstract}
\maketitle
\section{Introduction}
    \noindent
    Attempts to control and manipulate quantum systems for processing quantum information have inspired great advances in our understanding of quantum dynamics.
    For generic interacting quantum systems, time evolution scrambles quantum information on the approach to a thermal state, which limits our ability to efficiently process that information.
    The eigenstates of thermalizing systems satisfy the eigenstate thermalization hypothesis (ETH) \cite{Deutsch1991, Srednicki1994, D_Alessio_2016, Polkovnikov_2011} and encode information in exponentially complex observables, rendering them irretrievable.
    Exceptions to the ETH are therefore important for the encoding and manipulation of information in weakly entangled quantum states.
    In clean systems, the prototypical examples are the integrable systems, which possess an extensive number of integrals of motion \cite{Kinoshita2006, Schneider2012}, and scarred many-body Hamiltonians with a manifold of ETH-violating eigenstates \cite{Bernien2017a, Turner2018a, Moudgalya2018}.
    These both require an element of fine-tuning which leaves them unstable to perturbation.
    Instead, a promising avenue towards \emph{robustly} avoiding thermalization is through quenched disorder in a phenomenon known as many-body localization (MBL) \cite{Anderson1958, Basko2006, Huse2007, Pal2010, Imbrie2014, Nandkishore_2015, Abanin_2019, Schreiber2015, Choi2016}.
    The emergence of local integrals of motion (l-bits) in the MBL phase \cite{Serbyn2013, Huse2013b, Chandran2015b, Rademaker2016, Kulshreshtha2018, Goihl2018, Thomson2018} allows for the protection of \textit{classical} information effectively, however this performs poorly with quantum information \cite{Serbyn2014, Banuls2017}.

    The situation at zero temperature is a little different, where topological order and symmetry-protected topology (SPT) allow for the robust encoding of quantum information into a degenerate ground-state manifold \cite{Senthil_2015}.
    These are stable to (symmetry preserving) perturbations, provided the energy gap to the excited states does not close, but typically fail in the presence of thermal noise and delocalized excitations \cite{Pollman2010, Chen_2011, Fidkowski_2011} as the degeneracy in the ground states is usually not replicated in the highly-excited states.
    Strong disorder can stabilize topological degrees of freedom at finite energy density through MBL \cite{Bahri2013, Huse2013, Parameswaran2018}.
    Additionally, even in clean systems a strong zero mode (SZM) imposes a spectrum-wide energy pairing \cite{Kitaev2001, Fendley_2012, Fendley_2016} and can enable coherent storage of quantum information \cite{Kemp2019}.
    This has also been found recently in scarred Hamiltonians \cite{Srivatsa2020, Jeyaretnam2021}.

    The addition of disorder to a system with ground-state SPT order provides an avenue to stabilise that order at finite energy densities~\cite{Huse2013, Pekker2014, Kjall2014} -- producing a system with multiple topologically distinct MBL phases and possibly direct eigenstate-ordering phase transitions between them.
	However, recent exact diagonalization studies demonstrate that an ergodic phase may intervene at arbitrarily small interaction strengths, and some even claim that an MBL-to-MBL phase transition is forbidden \cite{Parameswaran2018, Sahay2021, Moudgalya2020b, Laflorencie2022}.
    Avalanches induced by rare regions \cite{Agarwal2017, Luitz2017, DeRoeck2017, Suntajs2020, Crowley2022b, Sels2022} and resonances \cite{Tikhonov2021, Morningstar2022, Long2022, Ha2023} play a crucial role in destabilizing MBL at the localization transition, and are candidates for generating the intervening delocalized phase here.

	Finite-depth tensor network techniques \cite{Pollman2016, Wahl2017, Pekker2017b, Wahl2022}, flow-equations \cite{Pekker2017, Thomson2018}, and renormalization group (RG) approaches \cite{Vosk2013, Vasseur2015, Goremykina2019} for approximating weakly entangled excited states provide access to dynamical properties, critical behavior, and l-bit operators, and have enriched our understanding of MBL.
	The RG techniques progressively eliminate or ``decimate'' degrees of freedom from a system, typically starting with the smallest length scales and highest frequencies, to arrive at a long-range or low-energy effective model \cite{Dasgupta1980, Fisher1992}.
	For disordered systems, we may make use of the real space RG for excited states (\rsrgx{}) \cite{Pekker2014, You2016}.
	The coarse-graining process lends itself to identifying the real-space structure of resonances, and studying their size and statistics.

	In this work we have applied \rsrgx{} to an interacting spin-$1/2$ chain with two MBL phases: a trivial paramagnetic phase and a spin glass phase with SPT order protected by a global $\mathbb{Z}_2$ symmetry.
	We have extracted a Clifford circuit and Schrieffer-Wolff transformation which together approximately diagonalize the Hamiltonian to first order \cite{You2016}.
	These encode the localized basis that would best fit the eigenstates of the Hamiltonian \textit{if} the system were localized, and we probe its stability to the off-diagonal part of the Hamiltonian by searching for many-body resonances \cite{Potter2015, DeRoeck2017b, Protopopov2017, Gornyi2017, Protopopov2020} between these basis states.
	Additionally, the geometry of these resonances, and in particular how these link l-bits together into thermal clusters, allows us to investigate the breakdown of localization through the use of a finite size scaling analysis.

	We find that the marginal MBL phase, as found in Ref.~\cite{Pekker2014}, is indeed destabilized to an ergodic phase for even relatively small interaction strengths, and that this phase may be extended in parameter space even with infinitesimal interactions.
	We show that the resonances filter through the system to form clusters that scale extensively with system size in the ergodic phase.
	We also look at the variance of the energy $\delta H^2$ of the \rsrgx{} basis, which quantifies the accuracy of these states as approximations to the true eigenstates.

	The rest of this paper proceeds as follows.
	In the next section, we lay out the details of the interacting Ising-Majorana model, and develop the Clifford \rsrgx{} technique as well as its application to this model.
	Then, in Sec.~\ref{sec:results}, we present the results of this work, including:
	the discovery of a strong edge zero mode in the localized SPT phase in Sec.\ref{sec:sg-order};
	the calculation of energy variance in Sec.~\ref{sec:energy_variance};
	a description of resonant mixing in the \rsrgx{} basis in Sec.~\ref{sec:resonances};
	the spatial distribution of these resonances in \ref{sec:resonant_clusters};
	and finally the scaling with increasing system size in Sec.~\ref{sec:resonance_scaling}, a key result of this paper.
	We finish in Sec.~\ref{sec:discussion} with a discussion of these results, giving our conclusions and suggesting future avenues of research.
\section{Model and RSRG-X}
	We consider a transverse field Ising model with nearest-neighbor and next-nearest-neighbor interactions, also known as the interacting Ising-Majorana model, described by the Hamiltonian
	\begin{equation}
		\label{eq:ham}
		H = \sum_{i} h_i \s{Z}_i + J_i \s{X}_i \s{X}_{i+1} + g \left( \s{Z}_i \s{Z}_{i+1} + \s{X}_i \s{X}_{i+2}\right)\ ,
	\end{equation}
	with $h_i \sim \text{Uniform}[0, h]$ and $J_i \sim \text{Uniform}[0, J]$.
	We normalize this by setting $hJ = 1$.
	We also use open boundary conditions.
	This model is statistically self-dual under the exchange $h \leftrightarrow J$, and is known to have two distinct MBL phases \cite{Huse2013, Pekker2014, Kjall2014}.
	On one hand, when $h$ is large, the local fields dominate and the model enters a topologically trivial paramagnetic (PM) phase, where the energy eigenstates are products states of frozen spins aligned along the $z$-axis.
	On the other hand, when $J$ is large, the model enters a spin glass (SG) phase, with spins forming large entangled clusters due to the action of the nearest-neighbor $\s{X}_i \s{X}_{i+1}$ terms.
	In this phase, the system is topologically ordered, protected by the global parity symmetry $G = \prod \s{Z}_i$, and hosts a Majorana edge zero mode \cite{Fendley_2012, Kemp_2017}.
	This zero mode leads to spectral pairing between the two parity sectors, with the gap exponentially small in system size \cite{Laflorencie2022}.
	We characterize the phase of the model by the quantity $\delta = \overline{\ln|J_i|} - \overline{\ln|h_i|} = 2\ln{J}$, such that the model is dual about $\delta \!=\! 0$ with positive and negative delta in the SG and PM phases respectively.

	In fact, if we choose to represent this Hamiltonian (via the Jordan-Wigner transformation) in terms of Majorana fermions such that two such fermions $(\gamma_{2i}, \gamma_{2i + 1})$ represent each physical spin $\sigma_i$, in an infinite or periodic chain the statistical duality above is made exact by regrouping the fermions into a new set of spins $\tilde{\sigma}_i$ each represented by $(\gamma_{2i - 1}, \gamma_{2i})$:
	\begin{equation}
		\begin{gathered}
			\includegraphics{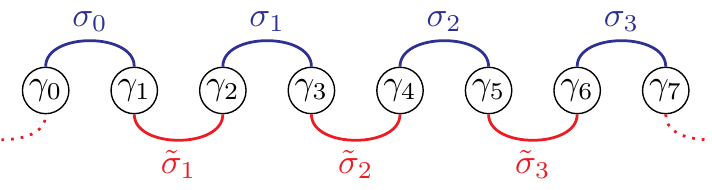}
		\end{gathered}
	\end{equation}

	Ground state RSRG has been used extensively to characterize similar models, including the model \eqref{eq:ham} in the fermionic representation \cite{Karcher2019}, but here we are interested in the excited states.
	Applied rigorously, \rsrgx{} relies on an assumption of strong  disorder, with relevant couplings distributed according to some power law $P(x) \propto x^\alpha$ (with $\alpha > 1$), leading to a good separation of energy scales in the system.
	In many cases the assumption of strong disorder may be relaxed: the system will quickly flow towards the infinite-randomness fixed point, and so the validity of the procedure is preserved.
	For all work in this paper, we set $g \ll \max(h, J)$, ensuring that one of the couplings $h_i$ and $J_i$ is always the largest in the system and thus the only couplings that need to be directly considered by the \rsrgx{} procedure.
	This keeps the Hamiltonian to a closed form.
	While the procedure can in principle generate a dominant interaction coupling, this is rare so long as $g$ is not too large, and we assume that this occurs infrequently enough not to meaningfully affect the disorder-averaged data.

	We therefore consider two types of decimation: the freezing of a single spin due to a dominant local field $h_i \s{Z}_i$ (``site decimation'') and the merger of two spins into one due to a dominant bond $J_i \s{X}_i \s{X}_{i+1}$ (``bond decimation'').
	In each case, the local Hilbert space is first rotated via an approximate Schrieffer–Wolff (SW) transformation \cite{Schrieffer1966}, truncated at second order, towards a basis aligned with the gap \cite{Pekker2014, You2016}, and then projected onto the subspace above or below this gap where the (transformed) local operator corresponding to the leading term ($\s{Z}_i$ or $\s{X}_i \s{X}_{i+1}$) is equal to $c = \pm 1$.
	The perturbative transformation removes terms that anticommute with the leading term, but also produces new terms which are second order in sub-leading energy scales.
	These new terms physically originate from the combination of two removed terms, mediated by the decimated spin: for example, when the term $h_2 \s{Z}_2$ is decimated, two nearest-neighbor terms $J_1 \s{X}_1 \s{X}_2$ and $J_2 \s{X}_2 \s{X}_3$ (which both anticommute with $h_2 \s{Z}_2$) are combined to form a term $c (J_1 J_2 / h_3) \s{X}_1 \s{X}_3$.
	At higher orders there can be contributions producing the same terms as those produced at second order and these interfering contributions are not included.
	The full RG rules may be found in Appendix~\ref{app:rsrg-x-decimation-rules}.

	\begin{figure}[tb]
		\includegraphics[width=\linewidth]{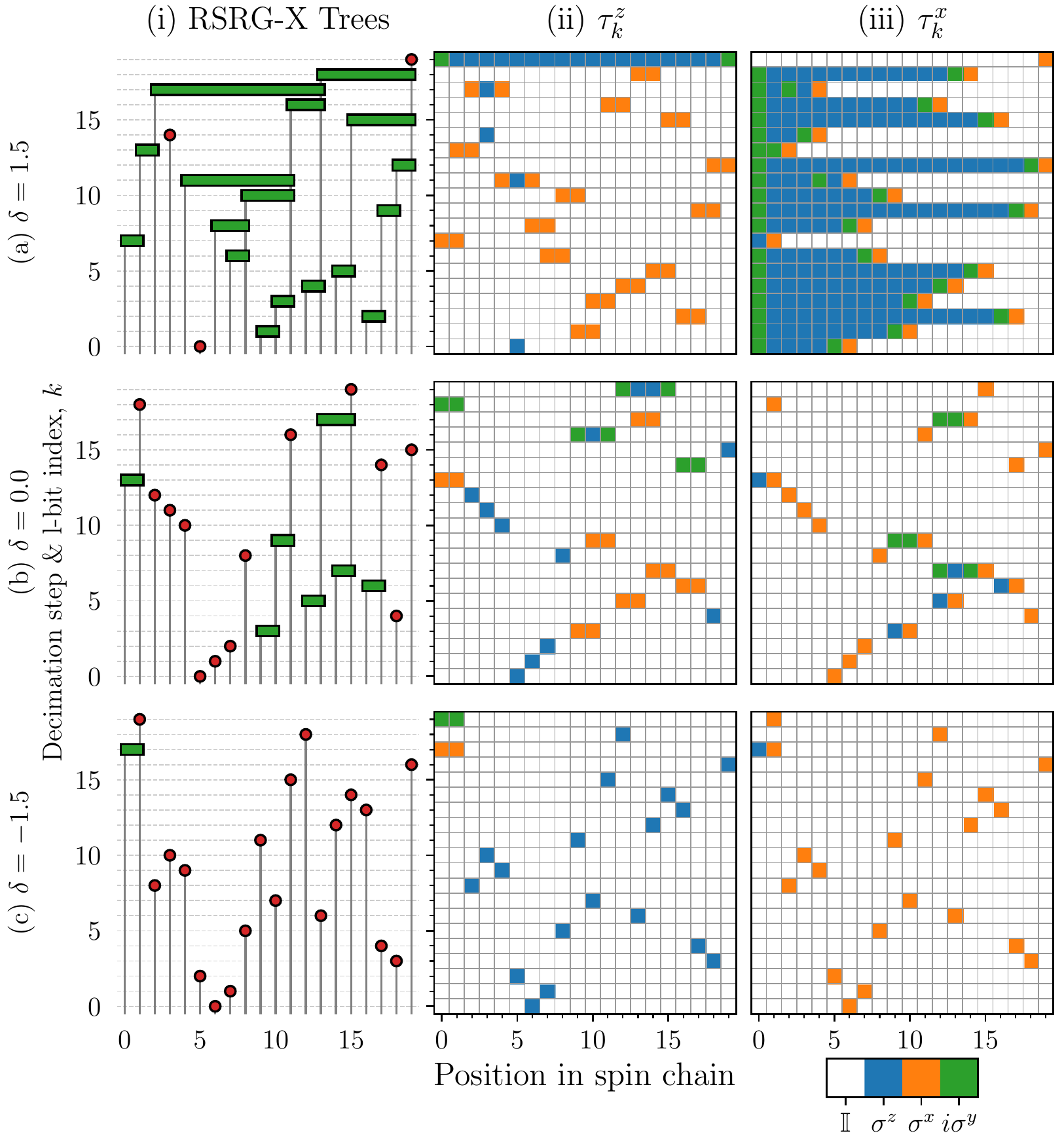}
		\caption{\label{fig:lbits}%
			For each of \textbf{(a)} the spin glass (SG) phase, $\delta=1.5$; \textbf{(b)} the critical phase, $\delta=0.0$; and \textbf{(c)}, the paramagnetic (PM) phase, $\delta=-1.5$, we generate a disorder realization for $L=20$ sites and apply \rsrgx{}.
			Additionally $g=0.2$ and we use open boundary conditions.
			The columns respectively show:
			\textbf{(i)} The tree tensor network corresponding to each state.
			Bond decimations are shown as green rectangles linking two effective spins into one;
			site decimations are red circles which freeze an effective spin.
			The y-axis shows the step in which each decimation occurred, corresponding also to a rough energy scale.
			\textbf{(ii)} The $z$-components of the l-bits $\left\{\tau_k\right\}$, which can also be seen as the stabilizers of this state.
			Each row, corresponding to a decimation on the left, shows a single l-bit, with the color giving the Pauli operator acting on each site.
			\textbf{(iii)} The $x$-components of the l-bits $\left\{\tau_{k}\right\}$, which can also be seen as the destabilizers.
		}
	\end{figure}
	We will later incorporate the SW transformations into our ansatz but if we neglect these, the successive merger of spins due to bond decimations causes the RG states to acquire a tree-like structure.
	As such these states can in fact be represented by tree tensor networks (TTNs) \cite{Protopopov2017, Protopopov2020, Ware2021}, where each node in the network represents a decimation, with $n$ incoming legs of bond dimension $d = 2$ for the spins to be decimated, and $n-1$ outgoing legs for the new effective spins after the decimation.
	The network takes us from $L$ physical spins and successively removes each degree of freedom, narrowing with each RG step.
	The tensor for a bond dimension is a projector from two spins onto one of the $\s{X}_i \s{X}_{i+1} = \pm 1$ subspaces, an isometry with two incoming legs and one outgoing leg.
	To represent spin up and down in the renormalized $\s{Z}$ basis, we choose respectively:
	\begin{align}
		\ket{a_c} &= \frac{1}{\sqrt{2}}\left(\ket{\uparrow\uparrow} + c \ket{\downarrow\downarrow}\right)\ ,\\
		\ket{b_c} &= \frac{1}{\sqrt{2}}\left(\ket{\uparrow\downarrow} - c \ket{\uparrow\downarrow}\right)\ .
	\end{align}
	The site decimation tensors are then projections of a single spin onto $\s{Z}_i = \pm 1$ -- just the two basis vectors in $d = 2$.
	This is represented pictorially in column~(i) of Fig.~\ref{fig:lbits}, where we show typical tree tensor networks for three points in the phase diagram: the spin glass phase, the paramagnetic phase, and the critical phase at $\delta \!=\! 0$.

	At each step we must choose between the excited and ground state manifolds, by selecting $c_k = \pm 1$.
	Since the energy shift of each decimation step typically decreases throughout the procedure, producing a hierarchy of energy scales, we can view the full set of possible choices as building up a branching tree of approximate eigenstates \cite{Pekker2014}.
	(This is not to be confused with the tree tensor network structure of the states themselves.)
	For this reason, we refer to a full set of choices and the corresponding approximate eigenstate as a ``leaf'' of the \rsrgx{} tree.
	The geometry of the TTN depends on the set of choices made for decimation directions $\{c_k\}$.
	However, if we choose to fix the geometry, we can re-interpret these TTNs by considering the choice of decimation direction $c_k = \pm 1$ as an additional outgoing leg.
	In this picture, degrees of freedom are not removed, but converted into the decimation choices $\{c_k\}$ which we then interpret as approximate l-bits.
	Hence, site decimations are the identity (since the resultant l-bit is exactly the Pauli $\s{Z}$ operator on that site), while bond decimations require us to map $\s{X}_i \s{X}_{i+1} \rightarrow \s{Z}_i$.
	Bearing in mind the Majorana duality between the PM and SG phases, we note that these operators are both fermion bilinears, and interpret the bond decimation as swapping $\gamma_{2i} \leftrightarrow \gamma_{2(i + 1)}$, which indeed achieves this mapping.
	This Majorana swap operation on adjacent spins may in fact be written as the following Clifford gate $R_b$,
	\begin{equation}\label{eq:bond-gate}
		\begin{gathered}
			\includegraphics{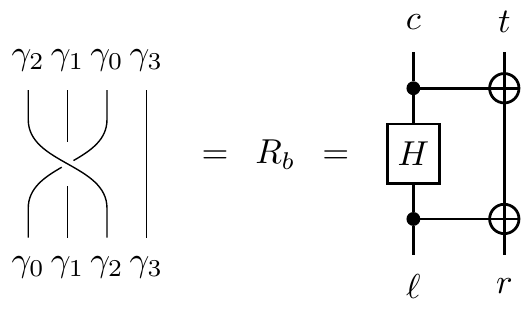}
		\end{gathered}
	\end{equation}

	where $\ell$ and $r = \ell + 1$ are the left- and right-hand spins, $t$ is the merged spin, and $c$ the decimation choice.
	This structure means that the operator mapping the spin basis onto the basis of decimation choices is a Clifford circuit, which we call $R$, and can be efficiently simulated \cite{Dehaene2003, Aaronson2004}.
	By applying the inverse transformation (from the l-bits to physical spins) to Pauli $\s{Z}$ and $\s{X}$ operators, we obtain the representation of the l-bits on the spin basis, $\{\tau_k^z\}$ and $\{\tau_k^x\}$.
	As an aside, these can also be viewed respectively as the stabilizers and destabilizers (see Refs.~\cite{Dehaene2003, Gottesman1997, Aaronson2004}) for the TTN states.

	In columns~(ii) and (iii) of Fig.~\ref{fig:lbits}, we show the $z$-\ and $x$-components respectively of the l-bits $\left\{\tau_k\right\}$ for the states corresponding to those in column (i).
	The PM phase in row~(c) contains mostly site decimations, where a single effective spin is frozen.
	This means that the l-bits are almost all single-site Pauli spins.
	On the other hand, the SG phase in row~(a) contains mostly bond decimations, which successively merge effective spins into large clusters represented by a tree-shaped network.
	A site decimation also freezes the final state of each tree.
	The stabilizers $\left\{\tau_k^z\right\}$ for these trees are two-site $\s{X}_\ell \s{X}_{\ell + 1}$ operators, and then the final site decimation is represented by a long operator $-\s{Y}_l \s{Z}_{l+1} \dots \s{Z}_{r - 1} \s{Y}_r$ across all the sites in the tree.
	The final site decimation in the SG phase typically has a very small energy scale associated with it, and may correspond to a strong zero mode linking the two symmetry sectors (see Sec.~\ref{sec:sg-order}).
	Finally, the critical phase in row~(b) is a mixture of these two phases, containing both PM and SG regions.
	There is some freedom in how we define the l-bits, and in particular we may multiply any l-bit $\tau_{k}^z$ by some other $\tau_{k'}^z$ (which it must commute with) to obtain other valid l-bits.

	The Clifford circuit representation allows us to efficiently calculate the action of any Pauli string (an operator that is the product of single-site Pauli operators) on a TTN state, by transforming it from the spin basis to the l-bit basis \cite{Dehaene2003, Aaronson2004}.
	Since the Clifford group maps Pauli strings to Pauli strings, this means such an operator maps one TTN state to exactly one other (with the same geometry), which may in fact be the same state.
	The Hamiltonian \eqref{eq:ham} is simply a sum of $O(L)$ Pauli strings which means that it retains this form in the l-bit basis, and so maps one TTN to at most $O(L)$ others.
	Therefore, we can efficiently calculate all matrix elements from a particular state, as well as expectation values.

	The geometry of these tree tensor networks is largely informed by the balance of local fields and bonds, quantified by the value of $\delta$.
	In order to better capture the effect of the interaction strength $g$, we also include the SW transformations in our wavefunction analysis.
	This is captured through an interaction picture: at each decimation the appropriate first-order SW transformation is calculated, $U_{\mathrm{SW}} = \exp\left(iS^{(1)}\right)$.
	We then apply the Clifford circuit, followed by the appropriately transformed SWs up to first order, to the Hamiltonian (or indeed to any operator) as $H^{(1)} = R^\dagger H R + \comm{iS^{(1)}}{R^\dagger H R}$.
	These SW transformations are analogous to the disentangling unitaries found in the multiscale entanglement renormalization ansatz (MERA)~\cite{Vidal2008}, which sit in between layers of isometries on a TTN.
	By connecting adjacent branches, these capture short-ranged entanglement locally and enable MERA to efficiently describe entanglement at all length scales.
	Applying the Clifford circuit and the SW transformations gives an effective Hamiltonian on the basis of l-bit product states, with those l-bits captured to first order.
	Each generator $S^{(1)}$ has support over a bounded number of sites, and so the effective Hamiltonian still has $O(L)$ terms.
	Despite this, computational complexity is still significantly increased, limiting the maximum system size accessible to the order of hundreds of spins rather than thousands for the ``zeroth-order'' calculations.
	For full details of the Clifford \rsrgx{} method, including the application of the SWs, see Appendix~\ref{app:clifford-RSRG-X}.

	The core output of our method therefore is a series of decimations, and associated SW transformations, which combine to create the localized state at a particular energy and also give us the effective local integrals of motion at that energy.
	These in turn are derived directly from the renormalized Hamiltonian at each step.
	Looking at the RG rules (App.~\ref{app:rsrg-x-decimation-rules}), one can see that the terms in this Hamiltonian generally have contributions from multiple sources, and so at a late stage in the process the coefficients encode the detailed history of decimation choices.
	One can view these rules as summing up different processes that interfere with each other, constructively or destructively, and as a result the dominant term at some RG step is conditioned on previous choices.
	This also means that the geometry of the TTNs is energy-dependent.

	This can be compared to the related spectrum bifurcation renormalization group (SBRG) technique \cite{You2016}.
	Unlike \rsrgx{}, SBRG aims to diagonalize the entire spectrum at once, extracting l-bit degrees of freedom but keeping these active (rotating them to $\s{z}\!$ operators) instead of removing them from the Hilbert space.
	This means crucially that the dominant term at any step only has contributions from  a single process, without interference, and the RG flow is not energy-dependent.
	Our method, in comparison, is therefore able to capture more detailed information about the states, at the cost of requiring us to select a particular target energy.
\section{Results}\label{sec:results}
	\subsection{Spin-glass order}\label{sec:sg-order}
		\begin{figure}
			\includegraphics[width=\linewidth]{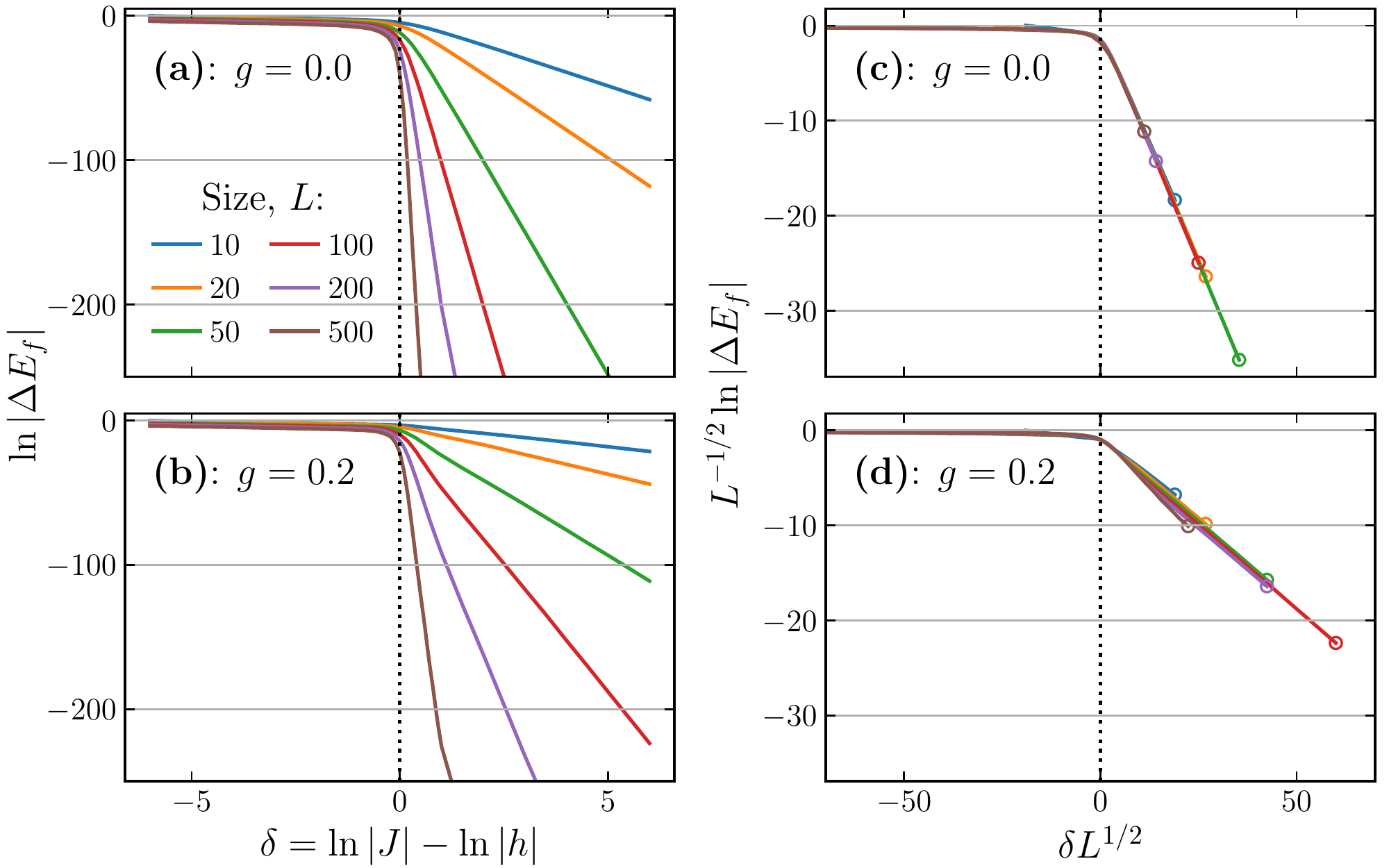}
			\caption{\label{fig:last_gap}%
				\textbf{(a, b)} Energy scale of the final l-bit $\Delta E_f$ produced during the \rsrgx{} procedure, as a function of $\delta$, for $g = 0$ and $g = 0.2$ respectively, and for various system sizes.
				\textbf{(c, d)} The same, but rescaled to show a clear data collapse and a phase transition at $\delta \!=\! 0$ for all interaction strengths.
				The data is truncated (at the hollow circles) where $\ln|\Delta E_f| < {-250}$, as this is beyond the limits of numerical precision.
				Note since \rsrgx{} assumes strong randomness, we cannot find evidence of an ergodic phase directly from calculations like this.
			}
		\end{figure}
		In Fig.~\ref{fig:last_gap}(a) and (b), we show the scaling of the final l-bit's energy $\Delta E_f$ (that is, the energy shift associated with the final \rsrgx{} decimation) with the parameter $\delta$, averaged across disorder realizations and decimation choices, for two different values of $g$ and a range of values of $L$.
		This should be the smallest energy scale in the system.
		Since the spin glass phase features spectral pairing \cite{Laflorencie2022}, we should expect this energy difference to be exponentially small in system size $L$ in the spin glass phase, but $O(1)$ in the paramagnetic phase.
		To test this, we plot $L^{-1/2}\ln|\Delta E_f|$ against $\delta L^{1/2}$ in panels (c) and (d).
		We observe a data collapse for both $g = 0$ and $g = 0.2$, although the collapse is much cleaner for the non-interacting case, with the line tending quickly to zero for $\delta < 0$ and to a straight line through the origin for positive $\delta > 0$.
		(It is possible therefore that this collapse is not universal, but that the exponents of $L$ here depend on the value of $g$.)
		Multiplying through by the factor of $L^{1/2}$, this clearly shows us that $\ln|\Delta E_f| \propto -L$ for fixed values of $g$ and $\delta > 0$, agreeing with predictions.
		The final l-bit in the SG phase, to leading order, also always takes the form of two $\s{Y}$ operators acting on either end of the largest spin cluster, with a string of $\s{Z}$ operators in between.
		Expressed in terms Majorana fermions, this is in fact a bilocalized operator acting on the two fermions at either ends of this string.
		Deep into the SG phase, the largest cluster spans the system, so this becomes an edge mode, reminiscent of those found in superconducting quantum wires \cite{Kitaev2001}.
		This operator anticommutes with the global parity operator $\prod_j \s{Z}_j$ and, given the exponentially small energy scale, this means that the final l-bit in the system in the SG phase is the strong zero mode (SZM) \cite{Fendley_2012, Fendley_2016, Kemp_2017}.
		However, in the presence of interactions or in the marginal regime, sub-leading corrections become significant.
		We could use the Schrieffer-Wolff transformation to find these, but leave this to future work.

        In applying \rsrgx{} we make an assumption of flow towards strong disorder, and the infinite-randomness fixed point -- implying the system is localized.
        Where we attempt to apply the method to a system which is in fact thermal, \rsrgx{} will produce inaccurate results.
        Previous work using exact, albeit small system size, numerical techniques \cite{Laflorencie2022, Moudgalya2020b, Protopopov2020}, suggests that such a thermal phase exists for $\delta \simeq 0$.
        Hence the collapse found above, which would ordinarily tell us that the transition at $\delta \!=\! 0$ becomes sharp in the thermodynamic limit, cannot be relied on as-is.
        What it \textit{does} imply is that, where we do in fact have localization and $\delta > 0$, we have an edge SZM.
        Therefore spectral pairing is always associated with the MBL SG phase, at least for the interaction strengths that we have looked at here.

		If we carefully probe the results from \rsrgx{} for accuracy, we can instead show by contradiction that the assumption of flow to strong disorder is violated, and therefore that we are in an ergodic phase.
		In particular, we can use Clifford \rsrgx{} to investigate the approximate eigenstates produced as leaves of the \rsrgx{} tree, and their stability to off-diagonal parts of the Hamiltonian.
	\subsection{Energy variance}\label{sec:energy_variance}
		\begin{figure}[tb]
			\includegraphics[width=\linewidth]{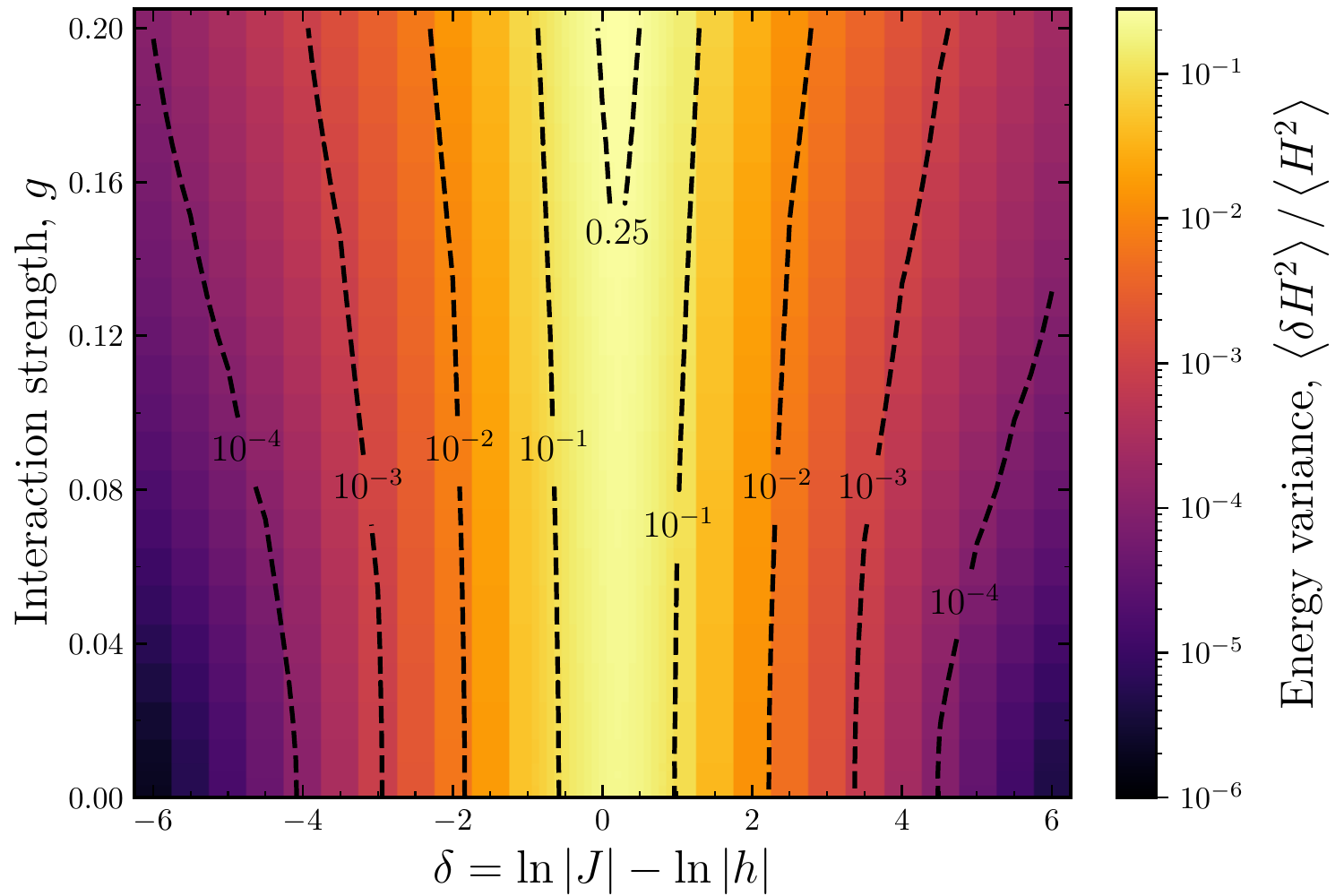}
			\caption{\label{fig:dH2}%
				Normalized energy variance $\expval{\delta H^2} / \expval{H^2}$ of \rsrgx{} leaf states across the $\delta$-$g$ plane, averaged over disorder realizations, for $L \!=\! 500$.
				A perfect eigenstate has zero energy variance, but this quantity increases as the quality of approximation worsens.
				Since \rsrgx{} leaf states are localized, a large energy variance may indicate delocalization and the approach to ergodicity.
				Also shown are selected contour lines (black).
			}
		\end{figure}
		For each disorder realization, we generate an \rsrgx{} leaf state $\ket{\psi_0}$ by picking a random set of decimation choices $\{c_k\}$ (that is, at infinite temperature), and then calculate the effective Hamiltonian on this state in the l-bit basis up to first order in SW transformations.
		We call this state the root state.
		This effective Hamiltonian is a sum of Pauli strings, with each Pauli string mapping the root state to exactly one state of definite l-bit configuration, $\ket{\psi_\alpha}$.
		We can then test the accuracy of the approximation by considering the ``energy variance'' \cite{You2016},
		\begin{equation}
			\label{eq:dH2}
			\expval{\delta H^2} = \matrixel{\psi_0}{H^2}{\psi_0} - |\!\matrixel{\psi_0}{H}{\psi_0}\!|^2\ ,
		\end{equation}
		which measures to what extent the root state $\ket{\psi_0}$ is a good approximation of an actual eigenstate of the Hamiltonian $H$.
		It can also be written as $\sum_{i \neq 0} |\!\matrixel{\psi_0}{H}{\psi_\alpha}\!|^2$, leading to an interpretation of $\expval{\delta H^2}$ as the degree to which the Hamiltonian maps the state away from itself.
		For perfect eigenstates, $\expval{\delta H^2} = 0$.

		Since $\expval{\delta H^2}$ scales with the square of the total energy, which grows with system size and is not consistent for different choices of parameters, we choose to normalize this quantity by the square of the Hamiltonian \textit{averaged across all states}, $\overline{\expval{H^2}} = \frac{1}{2^L} \Tr H^2$, which is basis independent.
		In Fig.~\ref{fig:dH2}, we plot this in the $\delta$-$g$ plane for a system of length $L\!=\!500$, showing that this quantity peaks towards the critical line $\delta\!=\!0$ and especially for larger values of $g$.
		One thing to note is that it appears to peak for small positive values of $\delta$, rather than exactly at $\delta\!=\!0$, and similar results are seen for many of the other quantities we calculate --
		this is due to the open boundary conditions which leave fewer (spin-glass) order terms in the Hamiltonian relative to (paramagnetic) disorder terms.

		The data show that the energy eigenstates in the critical region are less localized and cannot be well described by a single \rsrgx{} leaf state -- it is likely that as we approach the critical line, the eigenstates pick up large fluctuations, and hence multiple l-bit basis states are needed to capture them.
		This may imply delocalization of the eigenstates, but this depends on the details: if the number of basis states required grows extensively, then the system will thermalize in the thermodynamic limit, but otherwise these states will still occupy a vanishing fraction of the Hilbert space.
		To properly understand how the \rsrgx{} basis states hybridize to form the true eigenstates, we must therefore look at the resonances induced between them by off-diagonal Hamiltonian terms.
	\subsection{Many-body resonances}\label{sec:resonances}
		\begin{figure}[tb]
			\includegraphics[width=\linewidth]{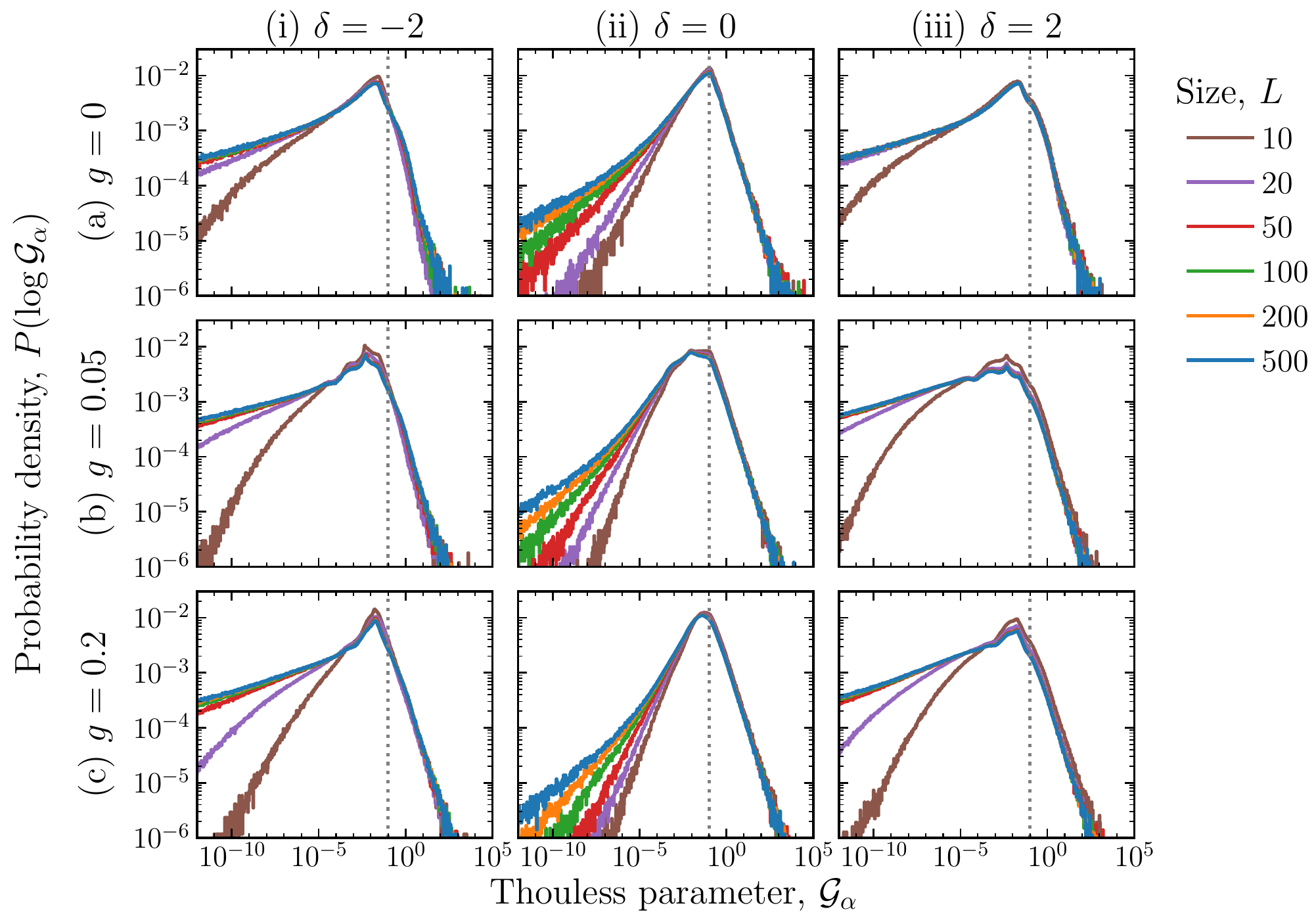}
			\caption{\label{fig:hist_G}%
				Probability density of $\ln(\mathcal{G}_\alpha)$ for various points in the $\delta$-$g$ plane, on a logarithmic scale.
				The columns show \textbf{(i)} the PM phase with $\delta = -2$, \textbf{(ii)} the critical regime with $\delta \!=\! 0$, and \textbf{(iii)} the SG phase with $\delta = 2$, while the rows show \textbf{(a)} the non-interacting case $g = 0$, \textbf{(b)} $g = 0.05$, and \textbf{(c)} $g = 0.2$.
				The distributions peak at large values of $\mathcal{G}_\alpha$ when $\delta \!=\! 0$, with more strongly decaying left-hand tails indicating that the weight of said distributions are much more concentrated at these large values.
				Note that we only include states $\ket{\psi_\alpha}$ with nonzero matrix elements to $\ket{\psi_0}$.
				The resonance threshold $\mathcal{G} = 0.1$ is indicated with a dashed vertical line.
			}
		\end{figure}
		Each term of the Hamiltonian maps the root state $\ket{\psi_0}$ to exactly one other tree tensor network state with the same geometry, which we label $\ket{\psi_\alpha}$.
		We can then calculate the many-body Thouless parameter \cite{Serbyn2015} for each term,
		\begin{equation}
			\label{eq:thouless}
			\mathcal{G}_\alpha = \left|\frac{\matrixel{\psi_0}{H}{\psi_\alpha}}{E_0 - E_\alpha}\right|\ .
		\end{equation}
		When $\mathcal{G}_\alpha \ll 1$, the Hamiltonian only weakly couples the root state to nearby states in the Hilbert space, implying that the true eigenstate is close to the root state with only small contributions from other states at low orders in perturbation theory.
		However, as this quantity grows larger, perturbation theory begins to break down, with the root state becoming strongly resonant with other nearby states.
		This implies that the true eigenstates are superpositions of multiple states in the computational basis.

		To make this a little more precise, for perturbation theory to converge the typical amplitude assigned to a diagram needs to decay faster than the combinatorial growth in the number of diagrams as the order of perturbation increases.
		We take $\mathcal{G}_\alpha$ as a rough estimate of this decay rate, and we choose to consider a resonance to have occurred when $\mathcal{G}_\alpha > \mathcal{G}^\ast = 0.1$.
		We believe this to be a slightly cautious threshold for what can be handled perturbatively.
		Any non-zero value of $\mathcal{G}_\alpha$ implies mixing of the tree tensor network basis; however, when small, we can take this to mean that the root state remains a good approximation of the true eigenstates up to some time, of order $1/\mathcal{G}^\ast$.
		Beyond that time perturbation theory would be required.
		In App.~\ref{app:higher_threshold} we give some data for the alternative choice $\mathcal{G}^\ast = 1$, which is a much more optimistic threshold for what can be handled perturbatively.

		The existence of a resonance does not necessarily mean that the eigenstates are no longer localized: when these resonances only connect a small number of states together (implying a small inverse participation ratio in the computational basis), the eigenstates may remain localized.
		Even if the number grows with system size, ergodicity is still avoided so long as the fraction of states involved vanishes in the thermodynamic limit.
		However, if resonances proliferate, they will connect an extensive number of states, leading to an ergodic phase.
		This is not the same as a thermal avalanche, wherein the resummation of one resonance creates a so-called superspin which, through the increased density of states, has an increased susceptibility to forming resonances with other l-bits.
		Here, we only consider the independent effects of off-diagonal terms and take clusters of directly resonant l-bits.

		We show the distribution of $\mathcal{G}_\alpha$ across disorder realizations and root states in Fig.~\ref{fig:hist_G}, for various values of $\delta$ and $g$.
		Close to the critical line $\delta \!=\! 0$ and with increasing interaction strength, these distributions peak more sharply (with faster-decaying left hand tails) at large values of $\mathcal{G}_\alpha$.
		Note that since each Hamiltonian term couples at most one other state to the root state, the total number of nonzero matrix elements is $O(L)$.
		Where two terms map to the same state, we combine their coefficients.

		\begin{figure}[tb]
			\includegraphics[width=\linewidth]{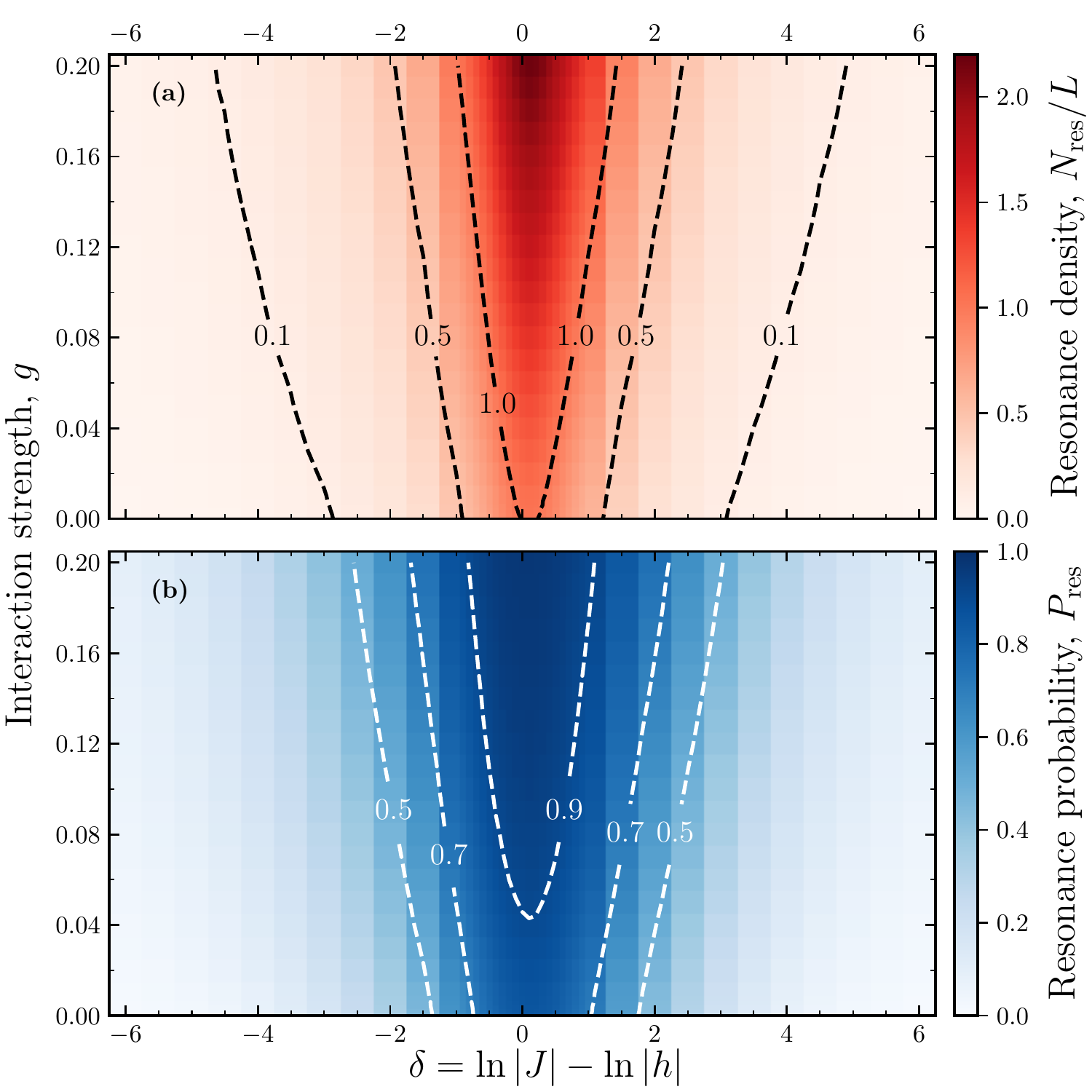}
			\caption{\label{fig:num_and_prob_res}%
				\textbf{(a)} Density (count per spin) of resonances with  $\mathcal{G}_\alpha > \mathcal{G}^\ast = 0.1$ and \textbf{(b)} probability that a particular l-bit in a random disorder realization is flipped by at least one such resonant term in the Hamiltonian.
				Both quantities calculated for randomly selected \rsrgx{} leaf states across the $\delta$-$g$ plane, averaged over disorder realizations, for $L \!=\! 500$.
				Also shown are selected contour lines (orange).
				While the precise phase boundary between localization and ergodicity cannot be located with this method, it is clear that the leaf states become unstable at increasingly small interaction strengths as one approaches the critical line $\delta \!=\! 0$, in agreement with the findings of Ref.~\cite{Laflorencie2022}.
			}
		\end{figure}
		In Fig.~\ref{fig:num_and_prob_res}(a), we count the number of terms for which the condition $\mathcal{G}_\alpha > \mathcal{G}^\ast$ is satisfied (that is, the number of resonances induced by the Hamiltonian), normalized by the size of the system.
		Towards the critical line $\delta \!=\! 0$ and at large (non-perturbative) interaction strengths $g$, we see this quantity peaking strongly, such that there is more than one resonance per spin in the system.
		This indicates a strong probability of crossover to an ergodic regime, although this does not allow us to locate the phase boundary precisely.

		We also consider the chance that a particular l-bit will be flipped by at least one resonance in Fig.~\ref{fig:num_and_prob_res}(b).
		This is subtly different to the average number of resonances per l-bit shown in (a) -- in that it is less strongly influenced by rare regions with large numbers of resonant terms affecting a small subset of the l-bits.
		This peaks at a small positive value of $\delta \!=\! 0$ and grows with increasing interaction strength $g$, to a greater than 90\% chance -- with almost every l-bit affected by a resonance, this makes it likely that the system is thermal in this portion of the phase diagram.
		To draw a more definitive conclusion however, we should look at the spatial distribution of these resonances.
	\subsection{Resonant Clusters}\label{sec:resonant_clusters}
		\begin{figure}[tb]
			\includegraphics[width=\linewidth]{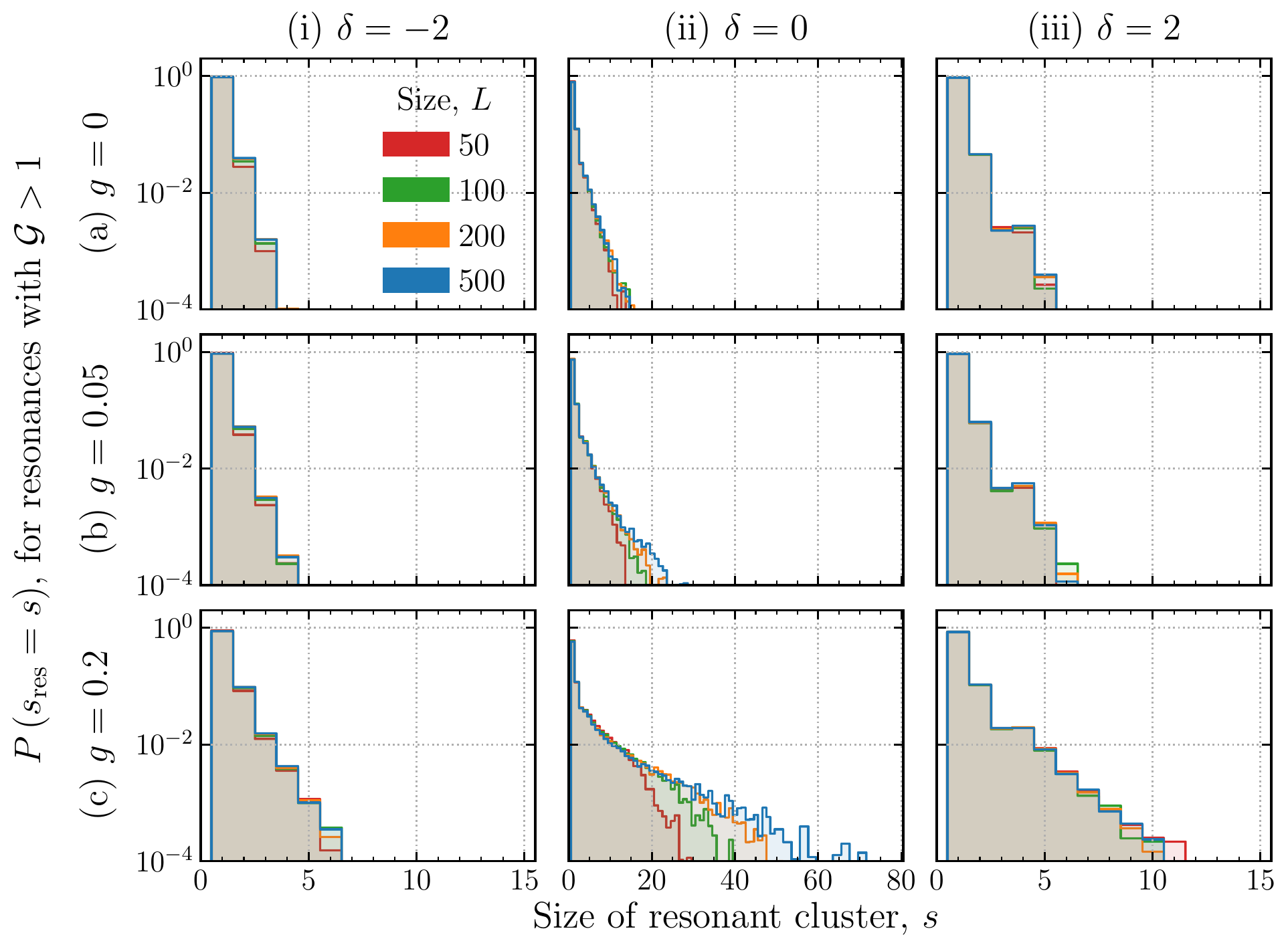}
			\caption{\label{fig:cluster_size_hists}%
				Probability that a particular l-bit is in a resonant cluster of size $s$, for nine choices of parameters: $g = 0, 0.05, 0.2$ respectively in each row, and for models in the spin-glass, critical, and paramagnetic phases respectively in each column.
				This should tend to a constant in the thermodynamic limit.
				For this figure, a value of $0$ indicates that an l-bit is unaffected by a resonant transition.
			}
		\end{figure}
		\begin{figure}[tb]
			\includegraphics[width=\linewidth]{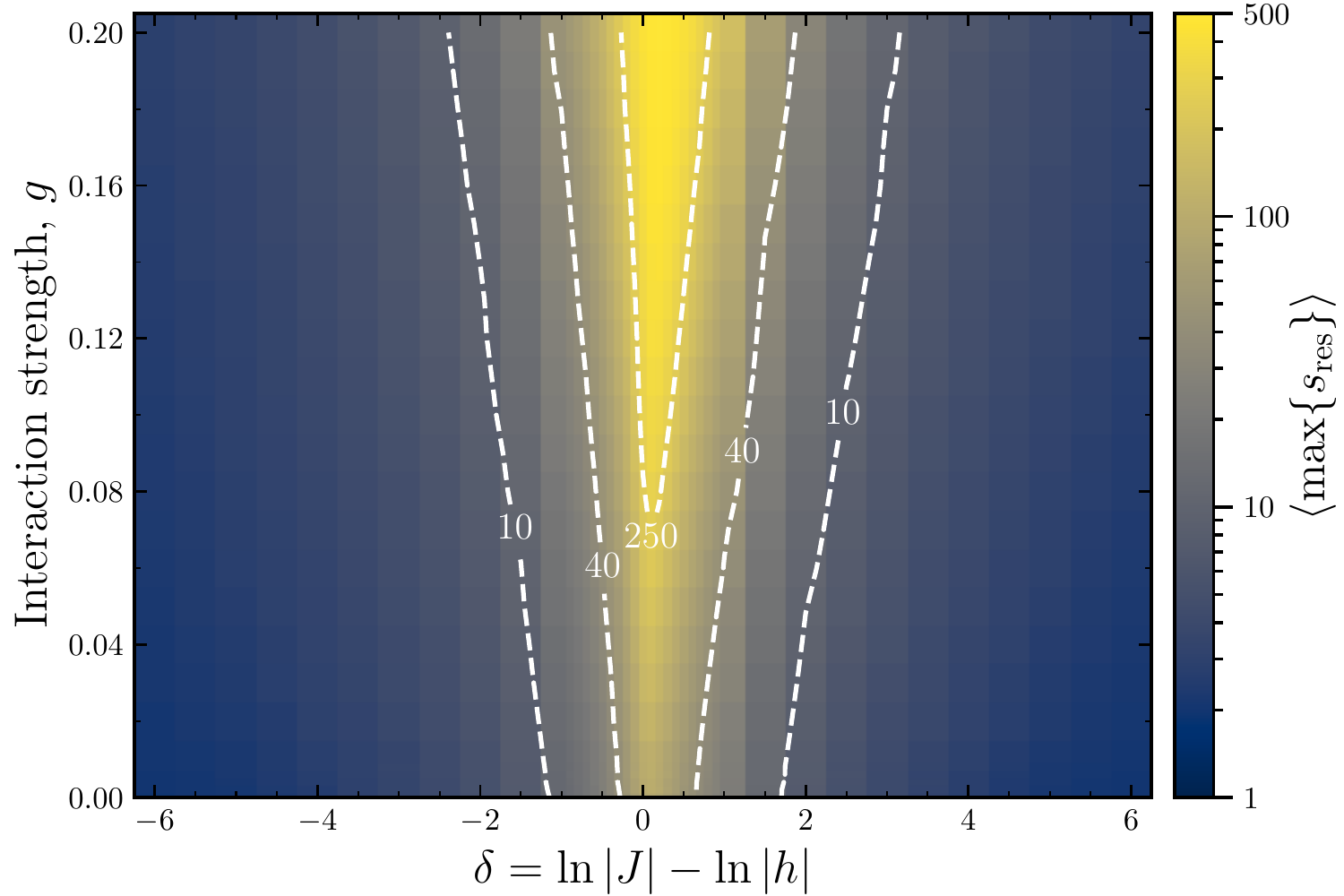}
			\caption{\label{fig:max_res_cluster_avg}%
				Maximum size of an l-bit cluster induced by resonances, averaged over disorder realizations, in the $\delta$-$g$ plane for $L \!=\! 500$.
				(This quantity is considered to be $1$ for a realization with no resonances).
				The data here is shown on a logarithmic scale.
				Also shown are selected contour lines (white).
			}
		\end{figure}
		Consider the sets of l-bits respectively flipped by each resonant term.
		By taking the union of those sets with non-zero intersection, one arrives at a natural definition of ``resonant clusters'': sets of degrees of freedom which are strongly mixed and locally thermal.
		This is analogous to percolation through a lattice where the links are formed by resonances.
		When resonances proliferate in a thermal phase \cite{Protopopov2017, Protopopov2020}, these clusters grow to occupy a significant fraction of the system size.

		In Fig.~\ref{fig:cluster_size_hists} we look at the distribution given by picking a random l-bit from a random disorder realization and calculating the number of l-bits in (the size of) the cluster it belongs to.
		(Here, an l-bit unaffected by resonances belongs to a cluster of size one.)
		When interactions are small and away from the critical line $\delta \!=\! 0$, the l-bits form small clusters, and the probability of a larger cluster forming rapidly tails off, exponentially with cluster size.
		However, as interactions get stronger and we move towards $\delta \!=\! 0$, the clusters grow larger, and in fact we can see these distributions are truncated by the finite system sizes accessible.
		For example, with $g = 0.2$ and $\delta \!=\! 0$, we see the distribution peaking at the size of the system -- it is overwhelmingly likely here that the resonances percolate through the entire system.

		Fig.~\ref{fig:max_res_cluster_avg} then shows the per-realization maximum size of these clusters, averaged over disorder, for a chain of length $L=500$.
		Note that this is length-dependent since longer systems give more opportunities for large clusters to form.
		This appears to peak for small but positive $\delta$, with the largest resonant clusters occupying almost the full system on average towards $g = 0.2$ and $\delta \!=\! 0$.
		However, even at small interaction strengths, the largest cluster typically spans a substantial fraction of the system.
		In App.~\ref{app:max_cluster}, we also show the probability distributions of the maximum cluster size over disorder realizations.

		The data support the hypothesis that at the critical line the l-bits strongly hybridize and move the system towards an ergodic phase, even for smaller interaction strengths.
		To get a complete picture, we need to understand the scaling behavior with system size. This will tell us whether resonances proliferate in the thermodynamic limit, indicating the breakdown of localization as the dynamics become ergodic; or whether resonances grow much slower than the system size, such that each eigenstate only occupies a vanishing fraction of the Hilbert space.
	\subsection{Scaling of resonance behavior with system size}\label{sec:resonance_scaling}
		\begin{figure}[tb]
			\includegraphics[width=\linewidth]{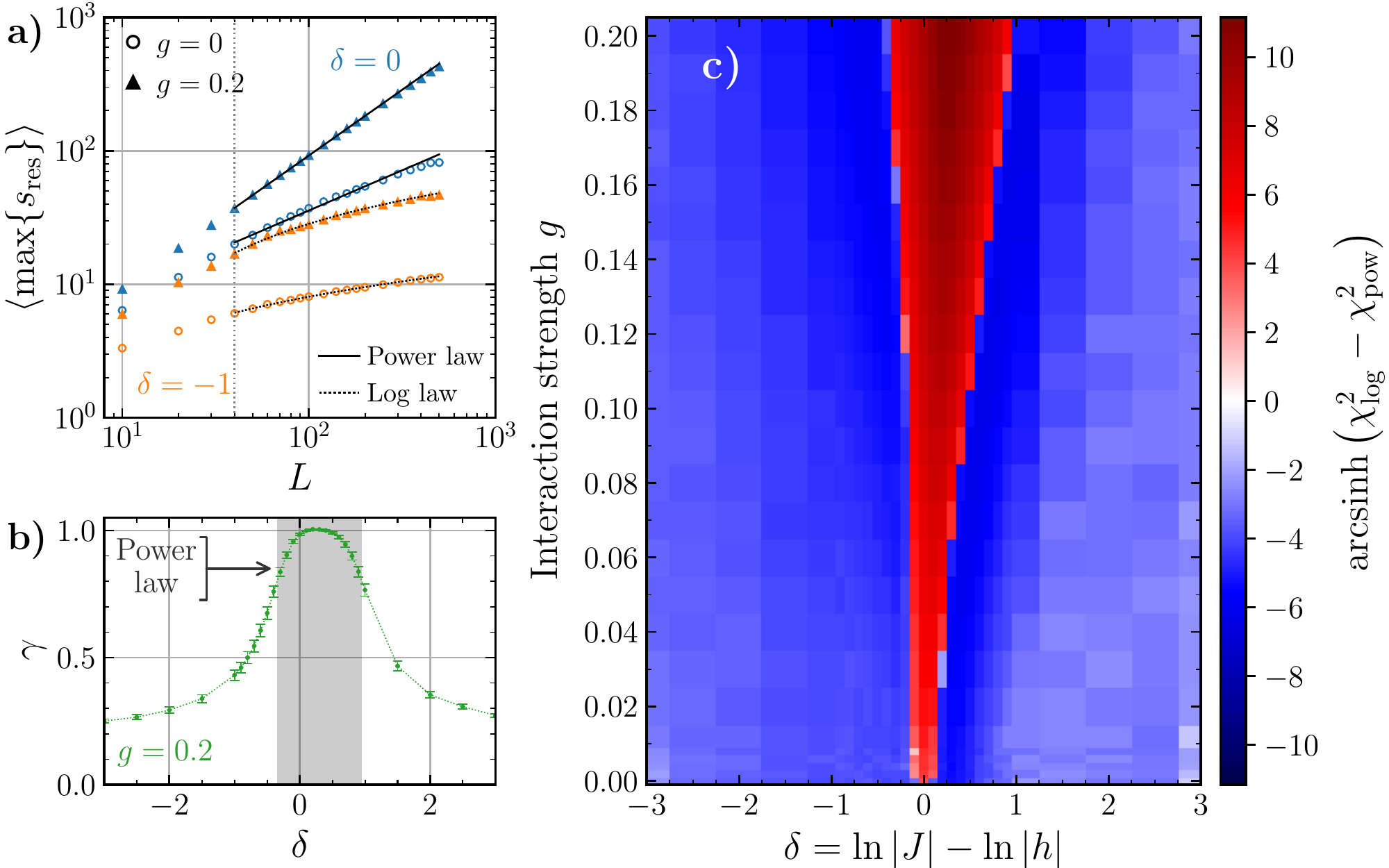}
			\caption{\label{fig:length_exponent}%
				\textbf{(a)}~Disorder-averaged maximal resonant cluster size $\left<\max\{s_\mathrm{res}\}\right>$ against system size $L$, for $\delta \!=\! 0$ (blue) and $\delta = -1$ (orange), each for $g = 0$ (open circles) and $g = 0.2$ (solid triangles).
                Solid and dotted lines indicate power-law ($\left<\max\{s_\mathrm{res}\}\right> \propto L^\gamma$) or log-law ($\propto \ln(L/L_0)$) fits respectively, to minimize the reduced-$\chi^2$ statistic.
				\textbf{(b)}~Fitted power-law exponent $\gamma$ against $\delta$ for fixed $g = 0.2$.
                The shaded region shows where the power-law fit is favored over a log-law fit.
                Here, $\gamma$ approaches 1, indicating the emergence of an extensive resonant cluster.
				\textbf{(c)}~Difference in the reduced-$\chi^2$ statistic between power-law and log-law fits, shown across the $\delta$--$g$ plane.
                For larger magnitude $\delta$, the dependence on $L$ grows weaker and the two hypotheses are more difficult to distinguish.
			}
		\end{figure}
		In order to uncover the behavior of the system in the thermodynamic limit, we need to understand how the resonant clusters scale with system size.
		To this end, in Fig.~\ref{fig:length_exponent}(a) we fit the mean maximal resonance size (see Figs.~\ref{fig:cluster_size_hists} and \ref{fig:max_res_cluster_avg}) to a power law in $L$, $\left<\max\{s_\mathrm{res}\}\right> \propto L^\gamma$, and to a log law, $\left<\max\{s_\mathrm{res}\}\right> \propto \ln(L/L_0)$.
		In the localized phase, where resonances are rare and well-separated spatially, we should expect cluster sizes to be exponentially distributed and therefore $\left<\max\{s_\mathrm{res}\}\right>$ ought to scale as a logarithm, following the maximum order statistic of an exponential distribution.
        By contrast in the ergodic phase, when resonances proliferate, we should expect the resultant clustering to percolate through the l-bits and form a cluster of size $O(L)$ such that $\gamma \rightarrow 1$.
		If the data follow a power law for smaller values of $\gamma$, this could still imply a power law in correlations and hence a diverging localization length, even if the largest cluster does not extend across the entire system.

		The data show that on the critical line $\delta \!=\! 0$, the size of the maximum cluster scales as a power law with $L$, with $\gamma \simeq 1.0$.
		This implies that resonances proliferate such that the largest cluster occupies a finite fraction of the Hilbert space in the thermodynamic limit, destabilizing the localized basis.
		Hence we conclude that the system is thermal at $\delta \!=\! 0$.
		Looking away from the critical line, in Fig.~\ref{fig:length_exponent}(b), we show the fitted power-law exponent $\gamma$ against $\delta$, for fixed interaction strength $g = 0.2$.
		Additionally, we shade the region in which a power law is favored over a log law.
		This shows that the power-law regime is accompanied by $\gamma \rightarrow 1$ and hence extensive resonant cluster scaling, verifying the intuition that this corresponds to a thermal phase.

		Finally, Fig.~\ref{fig:length_exponent}(c) gives the difference between the reduced-$\chi^2$ statistic between the power-law and log-law fits.
		The red region indicates that a power law is a better fit; the blue region corresponds to a log law.
		This shows clear evidence of an intervening ergodic phase, manifesting as a power law in resonant cluster scaling.
		The width of this phase grows with increasing $g$, but it is not entirely clear if this narrows to a single point for small but finite interaction strengths.
		(In our numerics, the smallest non-zero value we looked at was $g = 0.002$.)
\section{Discussion}\label{sec:discussion}
	In this work we have developed a method to apply real-space renormalization group to excited states (\rsrgx{}) of disordered spin-$1/2$ Hamiltonians and implicitly construct their wavefunctions as stabilizer states, even for large systems with many hundreds of spins, in the process also uncovering the l-bits for the system.
	This is done by constructing a Clifford circuit representing these approximate l-bits.
	Additionally, we have applied the Schrieffer-Wolff transformations to first order in order to improve accuracy, though at the cost of increased numerical complexity.

	We have then applied this Clifford \rsrgx{} to the interacting Ising-Majorana chain, a model known to host two distinct MBL phases, and investigated the crossover between localized and ergodic behavior in the supposed marginal MBL regime between those two phases.
	By calculating the many-body Thouless parameter giving the strength of perturbative mixing between basis states, we show that resonances proliferate in the marginal MBL regime.
	This is shown to result in an intervening ergodic phase with boundaries similar to those found in previous exact diagonalization studies \cite{Parameswaran2018, Sahay2021, Moudgalya2020b, Laflorencie2022}.

	Additionally, we have used Clifford \rsrgx{} to find the lowest-energy l-bit in the spin-glass phase. We show that this is a strong zero mode reminiscent of those found in superconducting quantum wires \cite{Kitaev2001}, with the leading term being a bilocalized Majorana fermion operator acting on either end of the largest spin cluster.
	Using this technique, it is possible to calculate higher-order corrections to this strong zero mode; however, we leave a systematic analysis to future work, instead focusing on the resonance picture and the ergodic regime in this work.
	Modifications of this technique may also allow access to higher-spin systems, enabling detailed characterization of their MBL phases through determination of the l-bits and their higher-order corrections.

	An ergodic phase has been argued to exist over an extended parameter regime even at arbitrarily weak interaction strength using ideas of a thermal avalanche, triggered by rare regions of weak disorder \cite{Laflorencie2022}.
	In contrast, we argue for an intervening thermal phase due a different mechanism unrelated to avalanches or rare regions.
	This is similar to the situation in small size numerics around the localization transition, where MBL is destabilized despite the low likelihood of rare regions \cite{Emmanouilidis2021}.
	Still, avalanches may yet produce a wider ergodic phase than we find here.
 	In the absence of a rigorous proof disallowing a direct MBL-MBL transition, it would be interesting to investigate other models of disorder such as stronger (power-law) disorder where this might occur.

    Another promising direction would be to turn to quasiperiodic systems where rare regions do not occur, thus cannot precipitate an avalanche, and correlations in the disorder could be tunable independently of the transition.
    These effects could potentially stabilize MBL, as has been suggested for arrays of superconducting qubits \cite{Varvelis2022}, or even lead to a direct MBL-MBL transition.
    There has also been recent work in constructing effective Hubbard models from continuous quasicrystalline models \cite{Gottlob2022}, and these may provide a more physically realistic testbed for these ideas than toy models such as the Aubry-Andre model.
\begin{acknowledgements}
	J.J.\ is supported by a UK Engineering \& Physical Sciences Research Council (EPSRC) studentship (Project Ref.~2252612) under the Doctoral Training Partnership with UCL (Grant Ref.~EP/R513143/1).
	C.J.T.\ is supported by an EPSRC fellowship (Grant Ref.~EP/W005743/1).
	A.P.\ is funded by the European Research Council (ERC) under the EU’s Horizon 2020 research and innovation program via Grant Agreement No.~853368.
	Statement of compliance with EPSRC policy framework on research data: This publication is theoretical work that does not require supporting research data.
	The authors acknowledge the use of the UCL Myriad High Performance Computing Facility (Myriad@UCL), and associated support services, in the completion of this work.
\end{acknowledgements}

\FloatBarrier
\appendix
\section{RSRG-X Decimation Rules}\label{app:rsrg-x-decimation-rules}
	In these equations we consider any coupling that crosses the open boundary of the system to be zero.
	Additionally let $J'_i$ and $K_i$ be the strength of the term $\s{Z}_i \s{Z}_{i+1}$ and $\s{X}_i \s{X}_{i+2}$ respectively.
	\paragraph{Site decimation rules} Suppose the largest gap is due to $h_3$.
	Then we decimate site 3, setting $\s{Z}_3 = c$, and renormalize the couplings as follows (with all unspecified couplings unaltered):
	\begin{gather}
		\tilde{h}_2 = h_2 + c J'_2\ ,\quad
		\tilde{h}_4 = h_4 + c J'_3\ ,\\
		\begin{gathered}
			\tilde{J}_1 = J_1 + c\frac{K_1 J_2}{h_3}\ ,\quad
			\tilde{J}_2 = K_2 + c\frac{J_2 J_3}{h_3}\ ,\\
			\tilde{J}_4 = J_4 + c\frac{K_3 J_3}{h_3}\ ,
		\end{gathered}\\
		\tilde{J'}_2 = 0\ ,\quad
		\tilde{K}_1 = c\frac{K_1 J_3}{h_3}\ ,\quad
		\tilde{K}_2 = c\frac{K_3 J_2}{h_3}\ .
	\end{gather}
	We also calculate the change in the energy as,
	\begin{equation}
		\Delta E = c \left(h_3 + \frac{J_2^2 + J_3^2 + K_1^2 + K_3^2}{2h_3}\right)\ .
	\end{equation}

	\paragraph{Bond decimation rules} Suppose the largest gap is due to $J_3$.
	Then we decimate the bond between sites $3$ and $4$, and the two sites are merged to create a new spin labeled $c$, renormalizing the couplings as follows (with all unspecified couplings unaltered):
	\begin{gather}
		\begin{gathered}
			\tilde{h}_2 = h_2 + c\frac{h_3 J'_2}{J_3}\ ,\quad
			\tilde{h}_5 = h_5 + c\frac{h_4 J'_4}{J_3}\ ,\\
			\tilde{h}_c = J'_3 + c\frac{h_3 h_4}{J_3}\ ,\\
		\end{gathered}\\
		\tilde{J}_2 = c J_2 + K_2\ , \quad
		\tilde{J}_c = J_4 + c K_3\ ,\\
		\tilde{J'}_2 = c\frac{h_4 J'_2}{J_3}\ ,\quad
		\tilde{J'}_c = c\frac{h_3 J'_4}{J_3}\ ,\\
		\tilde{K}_1 = c K_1, \quad
		\tilde{K}_2 = 0, \quad
		\tilde{K}_c = K_4.
	\end{gather}
	We also calculate the change in the energy as,
	\begin{equation}
		\Delta E = c \left(J_3 + \frac{h_3^2 + h_4^2 + J'^2_2 + J'^2_4}{2h_3}\right)\ .
	\end{equation}
\section{Clifford RSRG-X}\label{app:clifford-RSRG-X}
	The starting point of the Clifford \rsrgx{} method is to apply traditional \rsrgx{} to a system, as per Ref.~\cite{Pekker2014} -- specifically, to a system described by a Hamiltonian in which each term is a Pauli string (a product of single-site Pauli operators).
	\rsrgx{} starts with the full system of $L$ spins, and successively ``decimates'' degrees of freedom through the following prescription:
	\begin{enumerate}
		\item Locate the strongest term in the Hamiltonian $H_0~=~\lambda A$, responsible for the largest energy gap.
		\item Find and apply the Schrieffer-Wolff (SW) transformation $e^{iS}$ which transforms the Hamiltonian to commute with $H_0$ -- making the gap manifest.
		\item Apply a Clifford transformation $R$ to rotate $H_0$ to a Pauli $\s{Z}_{\ell}$ on some site $\ell$, and decimate that site by freezing $\s{Z}_{\ell} = \pm 1$.
		\item Return to step 1 and repeat until all degrees of freedom are frozen.
	\end{enumerate}

	In the process, at each step, two transformations are generated: one, a Clifford rotation which maps Pauli strings to Pauli strings, and two, a Schrieffer-Wolff transformation which is more complicated.
	In order to analyze the properties of wavefunctions (and other related features, such as the effective Hamiltonian on the localized basis and matrix elements of operators), in many cases it has been sufficient to only include the Clifford rotations.
	One can combine these to form a Clifford circuit which prepares the (approximate) localized basis from product states or, equivalently, maps operators on the physical spins to operators on the l-bits.
	Other RSRG-based methods have avoided additionally applying the SW transformation due to the increased complexity involved.

	However, this process only obtains the localized basis to zeroth-order: each l-bit, which acts as a stabilizer to the basis, is a single Pauli string in the computational basis.
	In this work we found that this was not sufficient to capture the variation due to the interaction terms, motivating us to include the SW transformations in order to compute the localized basis to higher order.
	The transformation $S$ at each step is given by the solution to,
	\begin{equation}
		\comm{e^{iS}(H_0 + V)e^{-iS}}{H_0} = 0\ ,
	\end{equation}
	where $H = H_0 + V$.
	This can be expanded out and solved order-by-order in $V$.
	In the case of a Hamiltonian expressed in terms of Pauli strings, with a leading term $H_0 = \lambda A$, this is solved to first order by $S^{(0)} = 0$ and,
	\begin{equation}
		S^{(1)} = \frac{1}{4i\lambda^2} \comm{H_0}{V}\ .
	\end{equation}
	Let us define $e^{iS_i}$ and $R_i$ to be the transformations at the $i^\mathrm{th}$ decimation step (here, we drop the superscript $(1)$).
	Then, we may write down the complete transformation which approximately diagonalizes the Hamiltonian as,
	\begin{equation}
		\begin{aligned}
					U &= (R_L e^{iS_L}) (R_{L-1} e^{iS_{L-1}}) \dots (R_2 e^{iS_2}) (R_1 e^{S_1})\\
					&= (e^{i\widetilde{S}_L} e^{i\widetilde{S}_{L-1}} \dots e^{i\widetilde{S}_1}) (R_L R_{L-1} \dots R_1)\\
					&\simeq e^{i\widetilde{S}} R\ .
		\end{aligned}
	\end{equation}
	Here, we have defined the notation,
	\begin{align}
		\widetilde{S}_i &= R_{[i, L]} S_i R_{[i, L]}^\dagger\ ,\\
		R_{[i, L]} &= R_{L} R_{L - 1} \dots R_i\ ,
	\end{align}
	such that $R_{[i, L]}$ is the partial Clifford circuit which transforms $S_i$ (defined on the effective spins at that RG step) so that it instead acts upon the l-bit basis.
	In this way, we separate the transformation into two parts: a Clifford circuit $R$, and a product of unitaries $e^{i\widetilde{S}}$.
	We are free here to treat the SW transformations as commuting, such that $e^A e^B = e^{A + B}$, since we are working to first order in $V$.
	We may then transform any operator (expressed as a sum of Pauli strings) to act upon the localized basis to first order by first pushing it through the Clifford circuit then applying the first-order SW transformation: $\hat{O} \rightarrow R^\dagger \hat{O} R +  \comm*{i\widetilde{S}}{R^\dagger \hat{O} R}$.
	This is still expressed as a sum of Pauli strings, and so calculation of matrix elements etc.\ on l-bit product states is straightforward -- each Pauli string maps a product state to exactly one state (perhaps itself).
	Correspondingly, we can calculate the first-order l-bits in the spin basis as $\tau^{x, z} = R\comm*{-i\widetilde{S}}{\s{x, z}}\! R^\dagger$.
	For the purpose of implementing these calculations, we make use of the formalism in Ref.~\cite{Dehaene2003}, representing Clifford circuits as binary matrices and Pauli strings as binary vectors (with an associated coefficient).
	\begin{figure}[t]
		\includegraphics[width=\linewidth]{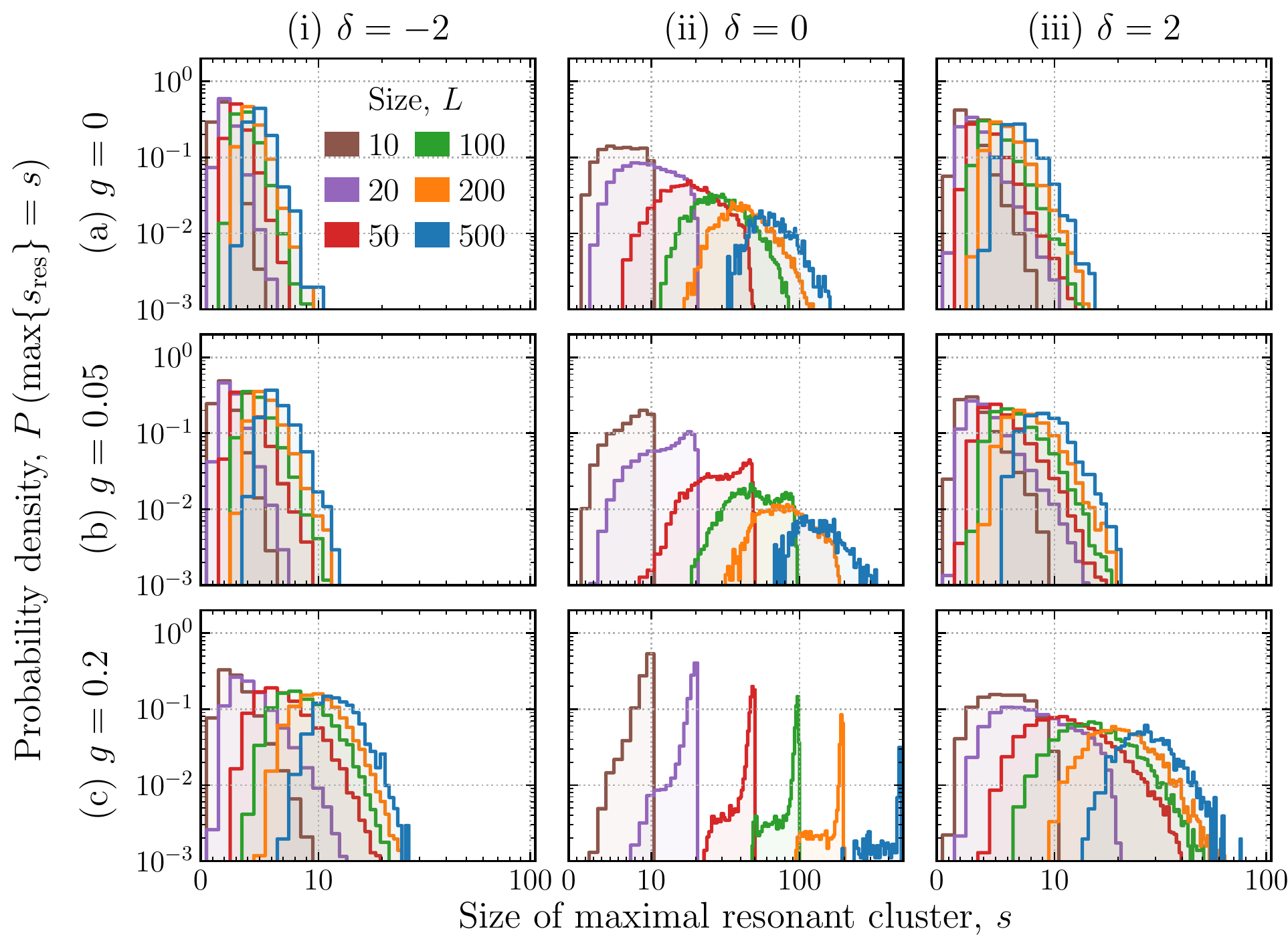}
		\caption{\label{fig:cluster_max_hists}%
			Distribution of maximal resonant l-bit cluster sizes for nine choices of parameters: $g = 0, 0.05, 0.2$ respectively in each row, for models in the spin-glass, critical, and paramagnetic phases respectively in each column. See Fig.~\ref{fig:max_res_cluster_avg} for comparison.
		}
	\end{figure}
\section{Distribution of maximal resonant cluster sizes}\label{app:max_cluster}
	In Fig.~\ref{fig:cluster_max_hists} we show histograms across disorder realizations of the maximum resonant cluster size, $\max\{s_\mathrm{res}\}$, for various system sizes $L$.
	The data show the maximum increasing with system size -- this is to be expected, as a larger system gives more chances for a large cluster to develop.
	Note also that in the thermal regime, the right-hand edge of the histograms is truncated by the system size with very little probability density on sizes smaller than this, showing that resonances dominate and are overwhelmingly likely to percolate throughout the entire system.
	\begin{figure}[tb]
		\includegraphics[width=\linewidth]{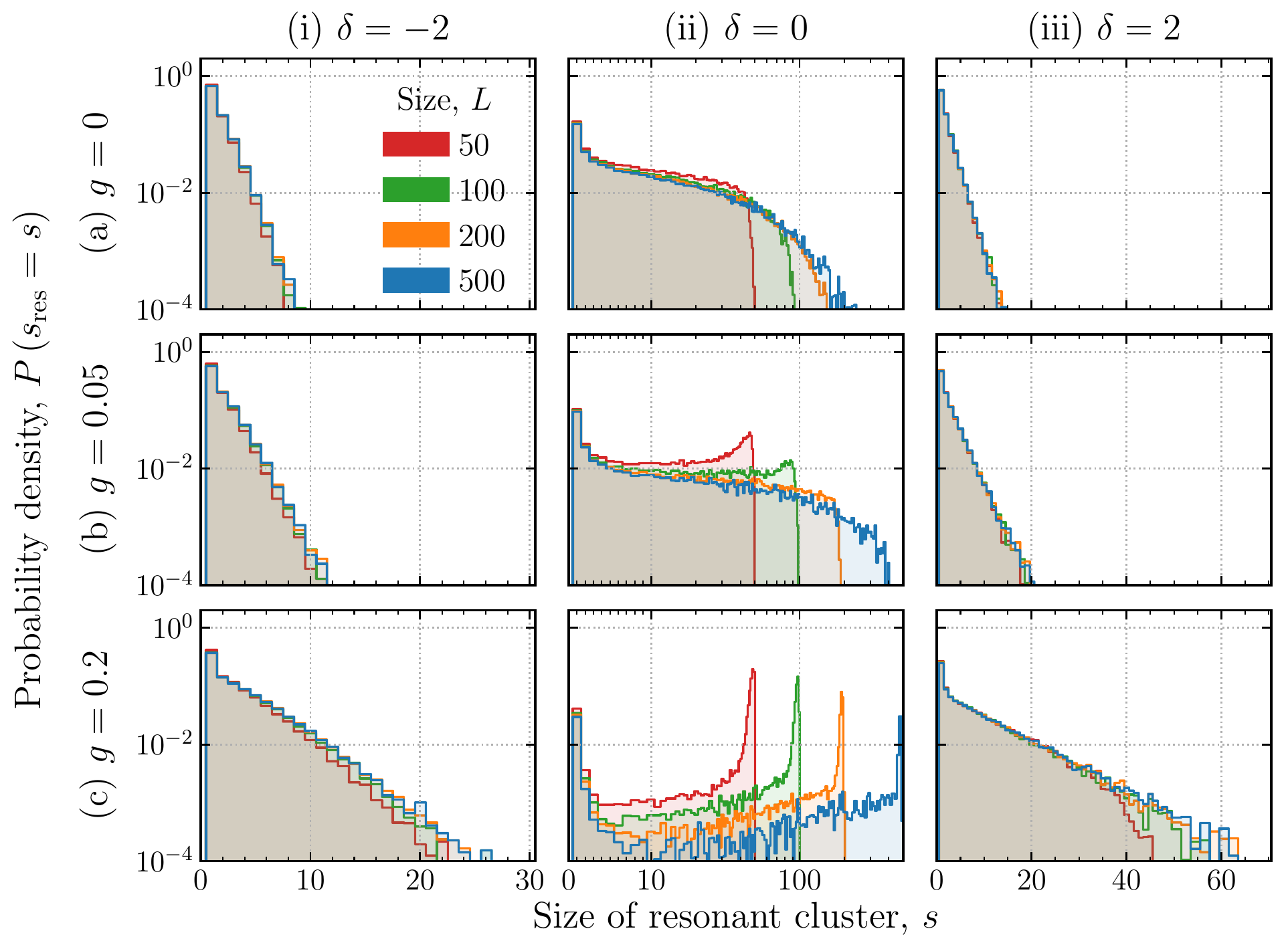}
		\caption{\label{fig:cluster_size_hists_large}%
			Similar to Fig.~\ref{fig:cluster_size_hists}, but only considering very strong resonances with $\mathcal{G} > 1$.
		}
	\end{figure}
\section{Higher resonance threshold}\label{app:higher_threshold}
	\begin{figure}[b]
		\includegraphics[width=\linewidth]{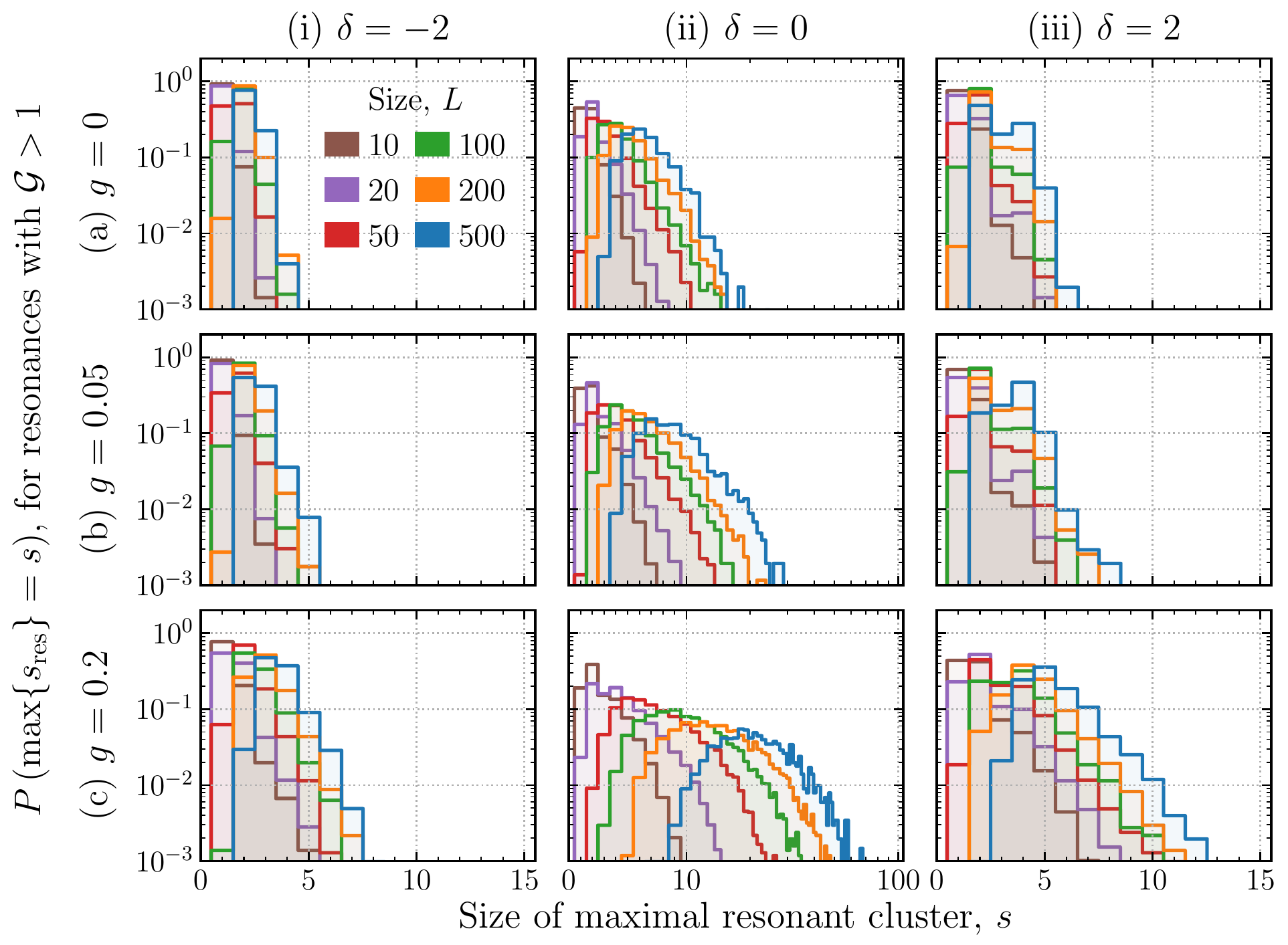}
		\caption{%
			Similar to Fig.~\ref{fig:cluster_max_hists}, but only considering very strong resonances with $\mathcal{G} > 1$.
		}
		\label{fig:cluster_max_hists_large}
	\end{figure}
	Throughout this work we have considered a resonance with $\mathcal{G} > \mathcal{G}^\ast = 0.1$ to be a resonance capable of destabilizing the localized basis.
	In this section, we show some data for a higher threshold, $\mathcal{G}^\ast = 1$, meaning that we only consider very strong resonances which are certain to destabilize the basis.
	Fig.~\ref{fig:cluster_size_hists_large} is an analogue of Fig.~\ref{fig:cluster_size_hists}, and shows the distribution of cluster sizes across disorder realizations.
	Additionally, Fig.~\ref{fig:cluster_max_hists_large} is an analogue of Fig.~\ref{fig:cluster_max_hists}, and shows the distribution of \textit{maximum} cluster sizes across disorder realizations.

	Despite taking a much more conservative estimate of what is necessary to destabilize the system, there is still a trend towards delocalization towards $\delta \simeq 0$ and $g \simeq 0.2$.
	The technique cannot (at present) be extended significantly beyond $g = 0.2$ because this would violate the assumption that the leading coupling comes from the relevant terms $h_i$ and $J_i$, but the trend is clear and it seems likely that the scaling of the maximum cluster size would become extensive for larger $g$.

\bibliography{RSRGX}

\begin{thebibliography}{83}%
\makeatletter
\providecommand \@ifxundefined [1]{%
 \@ifx{#1\undefined}
}%
\providecommand \@ifnum [1]{%
 \ifnum #1\expandafter \@firstoftwo
 \else \expandafter \@secondoftwo
 \fi
}%
\providecommand \@ifx [1]{%
 \ifx #1\expandafter \@firstoftwo
 \else \expandafter \@secondoftwo
 \fi
}%
\providecommand \natexlab [1]{#1}%
\providecommand \enquote  [1]{``#1''}%
\providecommand \bibnamefont  [1]{#1}%
\providecommand \bibfnamefont [1]{#1}%
\providecommand \citenamefont [1]{#1}%
\providecommand \href@noop [0]{\@secondoftwo}%
\providecommand \href [0]{\begingroup \@sanitize@url \@href}%
\providecommand \@href[1]{\@@startlink{#1}\@@href}%
\providecommand \@@href[1]{\endgroup#1\@@endlink}%
\providecommand \@sanitize@url [0]{\catcode `\\12\catcode `\$12\catcode
  `\&12\catcode `\#12\catcode `\^12\catcode `\_12\catcode `\%12\relax}%
\providecommand \@@startlink[1]{}%
\providecommand \@@endlink[0]{}%
\providecommand \url  [0]{\begingroup\@sanitize@url \@url }%
\providecommand \@url [1]{\endgroup\@href {#1}{\urlprefix }}%
\providecommand \urlprefix  [0]{URL }%
\providecommand \Eprint [0]{\href }%
\providecommand \doibase [0]{https://doi.org/}%
\providecommand \selectlanguage [0]{\@gobble}%
\providecommand \bibinfo  [0]{\@secondoftwo}%
\providecommand \bibfield  [0]{\@secondoftwo}%
\providecommand \translation [1]{[#1]}%
\providecommand \BibitemOpen [0]{}%
\providecommand \bibitemStop [0]{}%
\providecommand \bibitemNoStop [0]{.\EOS\space}%
\providecommand \EOS [0]{\spacefactor3000\relax}%
\providecommand \BibitemShut  [1]{\csname bibitem#1\endcsname}%
\let\auto@bib@innerbib\@empty
\bibitem [{\citenamefont {Deutsch}(1991)}]{Deutsch1991}%
  \BibitemOpen
  \bibfield  {author} {\bibinfo {author} {\bibfnamefont {J.~M.}\ \bibnamefont
  {Deutsch}},\ }\href {https://doi.org/10.1103/PhysRevA.43.2046} {\bibfield
  {journal} {\bibinfo  {journal} {Phys. Rev. A}\ }\textbf {\bibinfo {volume}
  {43}},\ \bibinfo {pages} {2046} (\bibinfo {year} {1991})}\BibitemShut
  {NoStop}%
\bibitem [{\citenamefont {Srednicki}(1994)}]{Srednicki1994}%
  \BibitemOpen
  \bibfield  {author} {\bibinfo {author} {\bibfnamefont {M.}~\bibnamefont
  {Srednicki}},\ }\href {https://doi.org/10.1103/PhysRevE.50.888} {\bibfield
  {journal} {\bibinfo  {journal} {Phys. Rev. E}\ }\textbf {\bibinfo {volume}
  {50}},\ \bibinfo {pages} {888} (\bibinfo {year} {1994})}\BibitemShut
  {NoStop}%
\bibitem [{\citenamefont {D'Alessio}\ \emph {et~al.}(2016)\citenamefont
  {D'Alessio}, \citenamefont {Kafri}, \citenamefont {Polkovnikov},\ and\
  \citenamefont {Rigol}}]{D_Alessio_2016}%
  \BibitemOpen
  \bibfield  {author} {\bibinfo {author} {\bibfnamefont {L.}~\bibnamefont
  {D'Alessio}}, \bibinfo {author} {\bibfnamefont {Y.}~\bibnamefont {Kafri}},
  \bibinfo {author} {\bibfnamefont {A.}~\bibnamefont {Polkovnikov}},\ and\
  \bibinfo {author} {\bibfnamefont {M.}~\bibnamefont {Rigol}},\ }\href
  {https://doi.org/10.1080/00018732.2016.1198134} {\bibfield  {journal}
  {\bibinfo  {journal} {Adv. Phys.}\ }\textbf {\bibinfo {volume} {65}},\
  \bibinfo {pages} {239} (\bibinfo {year} {2016})}\BibitemShut {NoStop}%
\bibitem [{\citenamefont {Polkovnikov}\ \emph {et~al.}(2011)\citenamefont
  {Polkovnikov}, \citenamefont {Sengupta}, \citenamefont {Silva},\ and\
  \citenamefont {Vengalattore}}]{Polkovnikov_2011}%
  \BibitemOpen
  \bibfield  {author} {\bibinfo {author} {\bibfnamefont {A.}~\bibnamefont
  {Polkovnikov}}, \bibinfo {author} {\bibfnamefont {K.}~\bibnamefont
  {Sengupta}}, \bibinfo {author} {\bibfnamefont {A.}~\bibnamefont {Silva}},\
  and\ \bibinfo {author} {\bibfnamefont {M.}~\bibnamefont {Vengalattore}},\
  }\href {https://doi.org/10.1103/RevModPhys.83.863} {\bibfield  {journal}
  {\bibinfo  {journal} {Rev. Mod. Phys.}\ }\textbf {\bibinfo {volume} {83}},\
  \bibinfo {pages} {863} (\bibinfo {year} {2011})}\BibitemShut {NoStop}%
\bibitem [{\citenamefont {Kinoshita}\ \emph {et~al.}(2006)\citenamefont
  {Kinoshita}, \citenamefont {Wenger},\ and\ \citenamefont
  {Weiss}}]{Kinoshita2006}%
  \BibitemOpen
  \bibfield  {author} {\bibinfo {author} {\bibfnamefont {T.}~\bibnamefont
  {Kinoshita}}, \bibinfo {author} {\bibfnamefont {T.}~\bibnamefont {Wenger}},\
  and\ \bibinfo {author} {\bibfnamefont {D.~S.}\ \bibnamefont {Weiss}},\ }\href
  {https://doi.org/10.1038/nature04693} {\bibfield  {journal} {\bibinfo
  {journal} {Nature}\ }\textbf {\bibinfo {volume} {440}},\ \bibinfo {pages}
  {900} (\bibinfo {year} {2006})}\BibitemShut {NoStop}%
\bibitem [{\citenamefont {Schneider}\ \emph {et~al.}(2012)\citenamefont
  {Schneider}, \citenamefont {Hackermüller}, \citenamefont {Ronzheimer},
  \citenamefont {Will}, \citenamefont {Braun}, \citenamefont {Best},
  \citenamefont {Bloch}, \citenamefont {Demler}, \citenamefont {Mandt},
  \citenamefont {Rasch},\ and\ \citenamefont {Rosch}}]{Schneider2012}%
  \BibitemOpen
  \bibfield  {author} {\bibinfo {author} {\bibfnamefont {U.}~\bibnamefont
  {Schneider}}, \bibinfo {author} {\bibfnamefont {L.}~\bibnamefont
  {Hackermüller}}, \bibinfo {author} {\bibfnamefont {J.~P.}\ \bibnamefont
  {Ronzheimer}}, \bibinfo {author} {\bibfnamefont {S.}~\bibnamefont {Will}},
  \bibinfo {author} {\bibfnamefont {S.}~\bibnamefont {Braun}}, \bibinfo
  {author} {\bibfnamefont {T.}~\bibnamefont {Best}}, \bibinfo {author}
  {\bibfnamefont {I.}~\bibnamefont {Bloch}}, \bibinfo {author} {\bibfnamefont
  {E.}~\bibnamefont {Demler}}, \bibinfo {author} {\bibfnamefont
  {S.}~\bibnamefont {Mandt}}, \bibinfo {author} {\bibfnamefont
  {D.}~\bibnamefont {Rasch}},\ and\ \bibinfo {author} {\bibfnamefont
  {A.}~\bibnamefont {Rosch}},\ }\href {https://doi.org/10.1038/nphys2205}
  {\bibfield  {journal} {\bibinfo  {journal} {Nat. Phys.}\ }\textbf {\bibinfo
  {volume} {8}},\ \bibinfo {pages} {213} (\bibinfo {year} {2012})}\BibitemShut
  {NoStop}%
\bibitem [{\citenamefont {Bernien}\ \emph {et~al.}(2017)\citenamefont
  {Bernien}, \citenamefont {Schwartz}, \citenamefont {Keesling}, \citenamefont
  {Levine}, \citenamefont {Omran}, \citenamefont {Pichler}, \citenamefont
  {Choi}, \citenamefont {Zibrov}, \citenamefont {Endres}, \citenamefont
  {Greiner}, \citenamefont {Vuletić},\ and\ \citenamefont
  {Lukin}}]{Bernien2017a}%
  \BibitemOpen
  \bibfield  {author} {\bibinfo {author} {\bibfnamefont {H.}~\bibnamefont
  {Bernien}}, \bibinfo {author} {\bibfnamefont {S.}~\bibnamefont {Schwartz}},
  \bibinfo {author} {\bibfnamefont {A.}~\bibnamefont {Keesling}}, \bibinfo
  {author} {\bibfnamefont {H.}~\bibnamefont {Levine}}, \bibinfo {author}
  {\bibfnamefont {A.}~\bibnamefont {Omran}}, \bibinfo {author} {\bibfnamefont
  {H.}~\bibnamefont {Pichler}}, \bibinfo {author} {\bibfnamefont
  {S.}~\bibnamefont {Choi}}, \bibinfo {author} {\bibfnamefont {A.~S.}\
  \bibnamefont {Zibrov}}, \bibinfo {author} {\bibfnamefont {M.}~\bibnamefont
  {Endres}}, \bibinfo {author} {\bibfnamefont {M.}~\bibnamefont {Greiner}},
  \bibinfo {author} {\bibfnamefont {V.}~\bibnamefont {Vuletić}},\ and\
  \bibinfo {author} {\bibfnamefont {M.~D.}\ \bibnamefont {Lukin}},\ }\href
  {https://doi.org/10.1038/nature24622} {\bibfield  {journal} {\bibinfo
  {journal} {Nature}\ }\textbf {\bibinfo {volume} {551}},\ \bibinfo {pages}
  {579} (\bibinfo {year} {2017})}\BibitemShut {NoStop}%
\bibitem [{\citenamefont {Turner}\ \emph {et~al.}(2018)\citenamefont {Turner},
  \citenamefont {Michailidis}, \citenamefont {Abanin}, \citenamefont {Serbyn},\
  and\ \citenamefont {Papi\'c}}]{Turner2018a}%
  \BibitemOpen
  \bibfield  {author} {\bibinfo {author} {\bibfnamefont {C.~J.}\ \bibnamefont
  {Turner}}, \bibinfo {author} {\bibfnamefont {A.~A.}\ \bibnamefont
  {Michailidis}}, \bibinfo {author} {\bibfnamefont {D.~A.}\ \bibnamefont
  {Abanin}}, \bibinfo {author} {\bibfnamefont {M.}~\bibnamefont {Serbyn}},\
  and\ \bibinfo {author} {\bibfnamefont {Z.}~\bibnamefont {Papi\'c}},\ }\href
  {https://doi.org/10.1038/s41567-018-0137-5} {\bibfield  {journal} {\bibinfo
  {journal} {Nat. Phys.}\ }\textbf {\bibinfo {volume} {14}},\ \bibinfo {pages}
  {745} (\bibinfo {year} {2018})}\BibitemShut {NoStop}%
\bibitem [{\citenamefont {Moudgalya}\ \emph {et~al.}(2018)\citenamefont
  {Moudgalya}, \citenamefont {Regnault},\ and\ \citenamefont
  {Bernevig}}]{Moudgalya2018}%
  \BibitemOpen
  \bibfield  {author} {\bibinfo {author} {\bibfnamefont {S.}~\bibnamefont
  {Moudgalya}}, \bibinfo {author} {\bibfnamefont {N.}~\bibnamefont
  {Regnault}},\ and\ \bibinfo {author} {\bibfnamefont {B.~A.}\ \bibnamefont
  {Bernevig}},\ }\href {https://doi.org/10.1103/PhysRevB.98.235156} {\bibfield
  {journal} {\bibinfo  {journal} {Phys. Rev. B}\ }\textbf {\bibinfo {volume}
  {98}},\ \bibinfo {pages} {235156} (\bibinfo {year} {2018})}\BibitemShut
  {NoStop}%
\bibitem [{\citenamefont {Anderson}(1958)}]{Anderson1958}%
  \BibitemOpen
  \bibfield  {author} {\bibinfo {author} {\bibfnamefont {P.~W.}\ \bibnamefont
  {Anderson}},\ }\href {https://doi.org/10.1103/PhysRev.109.1492} {\bibfield
  {journal} {\bibinfo  {journal} {Phys. Rev.}\ }\textbf {\bibinfo {volume}
  {109}},\ \bibinfo {pages} {1492} (\bibinfo {year} {1958})}\BibitemShut
  {NoStop}%
\bibitem [{\citenamefont {Basko}\ \emph {et~al.}(2006)\citenamefont {Basko},
  \citenamefont {Aleiner},\ and\ \citenamefont {Altshuler}}]{Basko2006}%
  \BibitemOpen
  \bibfield  {author} {\bibinfo {author} {\bibfnamefont {D.~M.}\ \bibnamefont
  {Basko}}, \bibinfo {author} {\bibfnamefont {I.~L.}\ \bibnamefont {Aleiner}},\
  and\ \bibinfo {author} {\bibfnamefont {B.~L.}\ \bibnamefont {Altshuler}},\
  }\href {https://doi.org/10.1016/j.aop.2005.11.014} {\bibfield  {journal}
  {\bibinfo  {journal} {Ann. Phys. (NY)}\ }\textbf {\bibinfo {volume} {321}},\
  \bibinfo {pages} {1126} (\bibinfo {year} {2006})}\BibitemShut {NoStop}%
\bibitem [{\citenamefont {Oganesyan}\ and\ \citenamefont
  {Huse}(2007)}]{Huse2007}%
  \BibitemOpen
  \bibfield  {author} {\bibinfo {author} {\bibfnamefont {V.}~\bibnamefont
  {Oganesyan}}\ and\ \bibinfo {author} {\bibfnamefont {D.~A.}\ \bibnamefont
  {Huse}},\ }\href {https://doi.org/10.1103/PhysRevB.75.155111} {\bibfield
  {journal} {\bibinfo  {journal} {Phys. Rev. B}\ }\textbf {\bibinfo {volume}
  {75}},\ \bibinfo {pages} {155111} (\bibinfo {year} {2007})}\BibitemShut
  {NoStop}%
\bibitem [{\citenamefont {Pal}\ and\ \citenamefont {Huse}(2010)}]{Pal2010}%
  \BibitemOpen
  \bibfield  {author} {\bibinfo {author} {\bibfnamefont {A.}~\bibnamefont
  {Pal}}\ and\ \bibinfo {author} {\bibfnamefont {D.~A.}\ \bibnamefont {Huse}},\
  }\href {https://doi.org/10.1103/PhysRevB.82.174411} {\bibfield  {journal}
  {\bibinfo  {journal} {Phys. Rev. B}\ }\textbf {\bibinfo {volume} {82}},\
  \bibinfo {pages} {174411} (\bibinfo {year} {2010})}\BibitemShut {NoStop}%
\bibitem [{\citenamefont {Imbrie}(2016)}]{Imbrie2014}%
  \BibitemOpen
  \bibfield  {author} {\bibinfo {author} {\bibfnamefont {J.~Z.}\ \bibnamefont
  {Imbrie}},\ }\href {https://doi.org/10.1007/s10955-016-1508-x} {\bibfield
  {journal} {\bibinfo  {journal} {J. Stat. Phys.}\ }\textbf {\bibinfo {volume}
  {163}},\ \bibinfo {pages} {998} (\bibinfo {year} {2016})}\BibitemShut
  {NoStop}%
\bibitem [{\citenamefont {Nandkishore}\ and\ \citenamefont
  {Huse}(2015)}]{Nandkishore_2015}%
  \BibitemOpen
  \bibfield  {author} {\bibinfo {author} {\bibfnamefont {R.}~\bibnamefont
  {Nandkishore}}\ and\ \bibinfo {author} {\bibfnamefont {D.~A.}\ \bibnamefont
  {Huse}},\ }\href {https://doi.org/10.1146/annurev-conmatphys-031214-014726}
  {\bibfield  {journal} {\bibinfo  {journal} {Annu. Rev. Condens. Matter
  Phys.}\ }\textbf {\bibinfo {volume} {6}},\ \bibinfo {pages} {15} (\bibinfo
  {year} {2015})}\BibitemShut {NoStop}%
\bibitem [{\citenamefont {Abanin}\ \emph {et~al.}(2019)\citenamefont {Abanin},
  \citenamefont {Altman}, \citenamefont {Bloch},\ and\ \citenamefont
  {Serbyn}}]{Abanin_2019}%
  \BibitemOpen
  \bibfield  {author} {\bibinfo {author} {\bibfnamefont {D.~A.}\ \bibnamefont
  {Abanin}}, \bibinfo {author} {\bibfnamefont {E.}~\bibnamefont {Altman}},
  \bibinfo {author} {\bibfnamefont {I.}~\bibnamefont {Bloch}},\ and\ \bibinfo
  {author} {\bibfnamefont {M.}~\bibnamefont {Serbyn}},\ }\href
  {https://doi.org/10.1103/RevModPhys.91.021001} {\bibfield  {journal}
  {\bibinfo  {journal} {Rev. Mod. Phys.}\ }\textbf {\bibinfo {volume} {91}},\
  \bibinfo {pages} {021001} (\bibinfo {year} {2019})}\BibitemShut {NoStop}%
\bibitem [{\citenamefont {Schreiber}\ \emph {et~al.}(2015)\citenamefont
  {Schreiber}, \citenamefont {Hodgman}, \citenamefont {Bordia}, \citenamefont
  {L\"uschen}, \citenamefont {Fischer}, \citenamefont {Vosk}, \citenamefont
  {Altman}, \citenamefont {Schneider},\ and\ \citenamefont
  {Bloch}}]{Schreiber2015}%
  \BibitemOpen
  \bibfield  {author} {\bibinfo {author} {\bibfnamefont {M.}~\bibnamefont
  {Schreiber}}, \bibinfo {author} {\bibfnamefont {S.~S.}\ \bibnamefont
  {Hodgman}}, \bibinfo {author} {\bibfnamefont {P.}~\bibnamefont {Bordia}},
  \bibinfo {author} {\bibfnamefont {H.~P.}\ \bibnamefont {L\"uschen}}, \bibinfo
  {author} {\bibfnamefont {M.~H.}\ \bibnamefont {Fischer}}, \bibinfo {author}
  {\bibfnamefont {R.}~\bibnamefont {Vosk}}, \bibinfo {author} {\bibfnamefont
  {E.}~\bibnamefont {Altman}}, \bibinfo {author} {\bibfnamefont
  {U.}~\bibnamefont {Schneider}},\ and\ \bibinfo {author} {\bibfnamefont
  {I.}~\bibnamefont {Bloch}},\ }\href {https://doi.org/10.1126/science.aaa7432}
  {\bibfield  {journal} {\bibinfo  {journal} {Science}\ }\textbf {\bibinfo
  {volume} {349}},\ \bibinfo {pages} {842} (\bibinfo {year}
  {2015})}\BibitemShut {NoStop}%
\bibitem [{\citenamefont {Choi}\ \emph {et~al.}(2016)\citenamefont {Choi},
  \citenamefont {Hild}, \citenamefont {Zeiher}, \citenamefont {Schauß},
  \citenamefont {Rubio-Abadal}, \citenamefont {Yefsah}, \citenamefont
  {Khemani}, \citenamefont {Huse}, \citenamefont {Bloch},\ and\ \citenamefont
  {Gross}}]{Choi2016}%
  \BibitemOpen
  \bibfield  {author} {\bibinfo {author} {\bibfnamefont {J.}~\bibnamefont
  {Choi}}, \bibinfo {author} {\bibfnamefont {S.}~\bibnamefont {Hild}}, \bibinfo
  {author} {\bibfnamefont {J.}~\bibnamefont {Zeiher}}, \bibinfo {author}
  {\bibfnamefont {P.}~\bibnamefont {Schauß}}, \bibinfo {author} {\bibfnamefont
  {A.}~\bibnamefont {Rubio-Abadal}}, \bibinfo {author} {\bibfnamefont
  {T.}~\bibnamefont {Yefsah}}, \bibinfo {author} {\bibfnamefont
  {V.}~\bibnamefont {Khemani}}, \bibinfo {author} {\bibfnamefont {D.~A.}\
  \bibnamefont {Huse}}, \bibinfo {author} {\bibfnamefont {I.}~\bibnamefont
  {Bloch}},\ and\ \bibinfo {author} {\bibfnamefont {C.}~\bibnamefont {Gross}},\
  }\href {https://doi.org/10.1126/science.aaf8834} {\bibfield  {journal}
  {\bibinfo  {journal} {Science}\ }\textbf {\bibinfo {volume} {352}},\ \bibinfo
  {pages} {1547} (\bibinfo {year} {2016})}\BibitemShut {NoStop}%
\bibitem [{\citenamefont {Serbyn}\ \emph {et~al.}(2013)\citenamefont {Serbyn},
  \citenamefont {Papi\'c},\ and\ \citenamefont {Abanin}}]{Serbyn2013}%
  \BibitemOpen
  \bibfield  {author} {\bibinfo {author} {\bibfnamefont {M.}~\bibnamefont
  {Serbyn}}, \bibinfo {author} {\bibfnamefont {Z.}~\bibnamefont {Papi\'c}},\
  and\ \bibinfo {author} {\bibfnamefont {D.~A.}\ \bibnamefont {Abanin}},\
  }\href {https://doi.org/10.1103/PhysRevLett.111.127201} {\bibfield  {journal}
  {\bibinfo  {journal} {Phys. Rev. Lett.}\ }\textbf {\bibinfo {volume} {111}},\
  \bibinfo {pages} {127201} (\bibinfo {year} {2013})}\BibitemShut {NoStop}%
\bibitem [{\citenamefont {Huse}\ \emph {et~al.}(2014)\citenamefont {Huse},
  \citenamefont {Nandkishore},\ and\ \citenamefont {Oganesyan}}]{Huse2013b}%
  \BibitemOpen
  \bibfield  {author} {\bibinfo {author} {\bibfnamefont {D.~A.}\ \bibnamefont
  {Huse}}, \bibinfo {author} {\bibfnamefont {R.}~\bibnamefont {Nandkishore}},\
  and\ \bibinfo {author} {\bibfnamefont {V.}~\bibnamefont {Oganesyan}},\ }\href
  {https://doi.org/10.1103/PhysRevB.90.174202} {\bibfield  {journal} {\bibinfo
  {journal} {Phys. Rev. B}\ }\textbf {\bibinfo {volume} {90}},\ \bibinfo
  {pages} {174202} (\bibinfo {year} {2014})}\BibitemShut {NoStop}%
\bibitem [{\citenamefont {Chandran}\ \emph {et~al.}(2015)\citenamefont
  {Chandran}, \citenamefont {Carrasquilla}, \citenamefont {Kim}, \citenamefont
  {Abanin},\ and\ \citenamefont {Vidal}}]{Chandran2015b}%
  \BibitemOpen
  \bibfield  {author} {\bibinfo {author} {\bibfnamefont {A.}~\bibnamefont
  {Chandran}}, \bibinfo {author} {\bibfnamefont {J.}~\bibnamefont
  {Carrasquilla}}, \bibinfo {author} {\bibfnamefont {I.~H.}\ \bibnamefont
  {Kim}}, \bibinfo {author} {\bibfnamefont {D.~A.}\ \bibnamefont {Abanin}},\
  and\ \bibinfo {author} {\bibfnamefont {G.}~\bibnamefont {Vidal}},\ }\href
  {https://doi.org/10.1103/PhysRevB.92.024201} {\bibfield  {journal} {\bibinfo
  {journal} {Phys. Rev. B}\ }\textbf {\bibinfo {volume} {92}},\ \bibinfo
  {pages} {024201} (\bibinfo {year} {2015})}\BibitemShut {NoStop}%
\bibitem [{\citenamefont {Rademaker}\ and\ \citenamefont
  {Ortu\~no}(2016)}]{Rademaker2016}%
  \BibitemOpen
  \bibfield  {author} {\bibinfo {author} {\bibfnamefont {L.}~\bibnamefont
  {Rademaker}}\ and\ \bibinfo {author} {\bibfnamefont {M.}~\bibnamefont
  {Ortu\~no}},\ }\href {https://doi.org/10.1103/PhysRevLett.116.010404}
  {\bibfield  {journal} {\bibinfo  {journal} {Phys. Rev. Lett.}\ }\textbf
  {\bibinfo {volume} {116}},\ \bibinfo {pages} {010404} (\bibinfo {year}
  {2016})}\BibitemShut {NoStop}%
\bibitem [{\citenamefont {Kulshreshtha}\ \emph {et~al.}(2018)\citenamefont
  {Kulshreshtha}, \citenamefont {Pal}, \citenamefont {Wahl},\ and\
  \citenamefont {Simon}}]{Kulshreshtha2018}%
  \BibitemOpen
  \bibfield  {author} {\bibinfo {author} {\bibfnamefont {A.~K.}\ \bibnamefont
  {Kulshreshtha}}, \bibinfo {author} {\bibfnamefont {A.}~\bibnamefont {Pal}},
  \bibinfo {author} {\bibfnamefont {T.~B.}\ \bibnamefont {Wahl}},\ and\
  \bibinfo {author} {\bibfnamefont {S.~H.}\ \bibnamefont {Simon}},\ }\href
  {https://doi.org/10.1103/PhysRevB.98.184201} {\bibfield  {journal} {\bibinfo
  {journal} {Phys. Rev. B}\ }\textbf {\bibinfo {volume} {98}},\ \bibinfo
  {pages} {184201} (\bibinfo {year} {2018})}\BibitemShut {NoStop}%
\bibitem [{\citenamefont {Goihl}\ \emph {et~al.}(2018)\citenamefont {Goihl},
  \citenamefont {Gluza}, \citenamefont {Krumnow},\ and\ \citenamefont
  {Eisert}}]{Goihl2018}%
  \BibitemOpen
  \bibfield  {author} {\bibinfo {author} {\bibfnamefont {M.}~\bibnamefont
  {Goihl}}, \bibinfo {author} {\bibfnamefont {M.}~\bibnamefont {Gluza}},
  \bibinfo {author} {\bibfnamefont {C.}~\bibnamefont {Krumnow}},\ and\ \bibinfo
  {author} {\bibfnamefont {J.}~\bibnamefont {Eisert}},\ }\href
  {https://doi.org/10.1103/PhysRevB.97.134202} {\bibfield  {journal} {\bibinfo
  {journal} {Phys. Rev. B}\ }\textbf {\bibinfo {volume} {97}},\ \bibinfo
  {pages} {134202} (\bibinfo {year} {2018})}\BibitemShut {NoStop}%
\bibitem [{\citenamefont {Thomson}\ and\ \citenamefont
  {Schir\'o}(2018)}]{Thomson2018}%
  \BibitemOpen
  \bibfield  {author} {\bibinfo {author} {\bibfnamefont {S.~J.}\ \bibnamefont
  {Thomson}}\ and\ \bibinfo {author} {\bibfnamefont {M.}~\bibnamefont
  {Schir\'o}},\ }\href {https://doi.org/10.1103/PhysRevB.97.060201} {\bibfield
  {journal} {\bibinfo  {journal} {Phys. Rev. B}\ }\textbf {\bibinfo {volume}
  {97}},\ \bibinfo {pages} {060201(R)} (\bibinfo {year} {2018})}\BibitemShut
  {NoStop}%
\bibitem [{\citenamefont {Serbyn}\ \emph {et~al.}(2014)\citenamefont {Serbyn},
  \citenamefont {Knap}, \citenamefont {Gopalakrishnan}, \citenamefont
  {Papi\'{c}}, \citenamefont {Yao}, \citenamefont {Laumann}, \citenamefont
  {Abanin}, \citenamefont {Lukin},\ and\ \citenamefont {Demler}}]{Serbyn2014}%
  \BibitemOpen
  \bibfield  {author} {\bibinfo {author} {\bibfnamefont {M.}~\bibnamefont
  {Serbyn}}, \bibinfo {author} {\bibfnamefont {M.}~\bibnamefont {Knap}},
  \bibinfo {author} {\bibfnamefont {S.}~\bibnamefont {Gopalakrishnan}},
  \bibinfo {author} {\bibfnamefont {Z.}~\bibnamefont {Papi\'{c}}}, \bibinfo
  {author} {\bibfnamefont {N.~Y.}\ \bibnamefont {Yao}}, \bibinfo {author}
  {\bibfnamefont {C.~R.}\ \bibnamefont {Laumann}}, \bibinfo {author}
  {\bibfnamefont {D.~A.}\ \bibnamefont {Abanin}}, \bibinfo {author}
  {\bibfnamefont {M.~D.}\ \bibnamefont {Lukin}},\ and\ \bibinfo {author}
  {\bibfnamefont {E.~A.}\ \bibnamefont {Demler}},\ }\href
  {https://doi.org/10.1103/PhysRevLett.113.147204} {\bibfield  {journal}
  {\bibinfo  {journal} {Phys. Rev. Lett.}\ }\textbf {\bibinfo {volume} {113}},\
  \bibinfo {pages} {147204} (\bibinfo {year} {2014})}\BibitemShut {NoStop}%
\bibitem [{\citenamefont {Ba\~nuls}\ \emph {et~al.}(2017)\citenamefont
  {Ba\~nuls}, \citenamefont {Yao}, \citenamefont {Choi}, \citenamefont
  {Lukin},\ and\ \citenamefont {Cirac}}]{Banuls2017}%
  \BibitemOpen
  \bibfield  {author} {\bibinfo {author} {\bibfnamefont {M.~C.}\ \bibnamefont
  {Ba\~nuls}}, \bibinfo {author} {\bibfnamefont {N.~Y.}\ \bibnamefont {Yao}},
  \bibinfo {author} {\bibfnamefont {S.}~\bibnamefont {Choi}}, \bibinfo {author}
  {\bibfnamefont {M.~D.}\ \bibnamefont {Lukin}},\ and\ \bibinfo {author}
  {\bibfnamefont {J.~I.}\ \bibnamefont {Cirac}},\ }\href
  {https://doi.org/10.1103/PhysRevB.96.174201} {\bibfield  {journal} {\bibinfo
  {journal} {Phys. Rev. B}\ }\textbf {\bibinfo {volume} {96}},\ \bibinfo
  {pages} {174201} (\bibinfo {year} {2017})}\BibitemShut {NoStop}%
\bibitem [{\citenamefont {Senthil}(2015)}]{Senthil_2015}%
  \BibitemOpen
  \bibfield  {author} {\bibinfo {author} {\bibfnamefont {T.}~\bibnamefont
  {Senthil}},\ }\href
  {https://doi.org/10.1146/annurev-conmatphys-031214-014740} {\bibfield
  {journal} {\bibinfo  {journal} {Annu. Rev. Condens. Matter Phys.}\ }\textbf
  {\bibinfo {volume} {6}},\ \bibinfo {pages} {299} (\bibinfo {year}
  {2015})}\BibitemShut {NoStop}%
\bibitem [{\citenamefont {Pollmann}\ \emph {et~al.}(2010)\citenamefont
  {Pollmann}, \citenamefont {Turner}, \citenamefont {Berg},\ and\ \citenamefont
  {Oshikawa}}]{Pollman2010}%
  \BibitemOpen
  \bibfield  {author} {\bibinfo {author} {\bibfnamefont {F.}~\bibnamefont
  {Pollmann}}, \bibinfo {author} {\bibfnamefont {A.~M.}\ \bibnamefont
  {Turner}}, \bibinfo {author} {\bibfnamefont {E.}~\bibnamefont {Berg}},\ and\
  \bibinfo {author} {\bibfnamefont {M.}~\bibnamefont {Oshikawa}},\ }\href
  {https://doi.org/10.1103/PhysRevB.81.064439} {\bibfield  {journal} {\bibinfo
  {journal} {Phys. Rev. B}\ }\textbf {\bibinfo {volume} {81}},\ \bibinfo
  {pages} {064439} (\bibinfo {year} {2010})}\BibitemShut {NoStop}%
\bibitem [{\citenamefont {Chen}\ \emph {et~al.}(2011)\citenamefont {Chen},
  \citenamefont {Gu},\ and\ \citenamefont {Wen}}]{Chen_2011}%
  \BibitemOpen
  \bibfield  {author} {\bibinfo {author} {\bibfnamefont {X.}~\bibnamefont
  {Chen}}, \bibinfo {author} {\bibfnamefont {Z.-C.}\ \bibnamefont {Gu}},\ and\
  \bibinfo {author} {\bibfnamefont {X.-G.}\ \bibnamefont {Wen}},\ }\href
  {https://doi.org/10.1103/PhysRevB.83.035107} {\bibfield  {journal} {\bibinfo
  {journal} {Phys. Rev. B}\ }\textbf {\bibinfo {volume} {83}},\ \bibinfo
  {pages} {035107} (\bibinfo {year} {2011})}\BibitemShut {NoStop}%
\bibitem [{\citenamefont {Fidkowski}\ and\ \citenamefont
  {Kitaev}(2011)}]{Fidkowski_2011}%
  \BibitemOpen
  \bibfield  {author} {\bibinfo {author} {\bibfnamefont {L.}~\bibnamefont
  {Fidkowski}}\ and\ \bibinfo {author} {\bibfnamefont {A.}~\bibnamefont
  {Kitaev}},\ }\href {https://doi.org/10.1103/PhysRevB.83.075103} {\bibfield
  {journal} {\bibinfo  {journal} {Phys. Rev. B}\ }\textbf {\bibinfo {volume}
  {83}},\ \bibinfo {pages} {075103} (\bibinfo {year} {2011})}\BibitemShut
  {NoStop}%
\bibitem [{\citenamefont {Bahri}\ \emph {et~al.}(2015)\citenamefont {Bahri},
  \citenamefont {Vosk}, \citenamefont {Altman},\ and\ \citenamefont
  {Vishwanath}}]{Bahri2013}%
  \BibitemOpen
  \bibfield  {author} {\bibinfo {author} {\bibfnamefont {Y.}~\bibnamefont
  {Bahri}}, \bibinfo {author} {\bibfnamefont {R.}~\bibnamefont {Vosk}},
  \bibinfo {author} {\bibfnamefont {E.}~\bibnamefont {Altman}},\ and\ \bibinfo
  {author} {\bibfnamefont {A.}~\bibnamefont {Vishwanath}},\ }\href
  {https://doi.org/10.1038/ncomms8341} {\bibfield  {journal} {\bibinfo
  {journal} {Nat. Commun.}\ }\textbf {\bibinfo {volume} {6}},\ \bibinfo {pages}
  {7341} (\bibinfo {year} {2015})}\BibitemShut {NoStop}%
\bibitem [{\citenamefont {Huse}\ \emph {et~al.}(2013)\citenamefont {Huse},
  \citenamefont {Nandkishore}, \citenamefont {Oganesyan}, \citenamefont {Pal},\
  and\ \citenamefont {Sondhi}}]{Huse2013}%
  \BibitemOpen
  \bibfield  {author} {\bibinfo {author} {\bibfnamefont {D.~A.}\ \bibnamefont
  {Huse}}, \bibinfo {author} {\bibfnamefont {R.}~\bibnamefont {Nandkishore}},
  \bibinfo {author} {\bibfnamefont {V.}~\bibnamefont {Oganesyan}}, \bibinfo
  {author} {\bibfnamefont {A.}~\bibnamefont {Pal}},\ and\ \bibinfo {author}
  {\bibfnamefont {S.~L.}\ \bibnamefont {Sondhi}},\ }\href
  {https://doi.org/10.1103/PhysRevB.88.014206} {\bibfield  {journal} {\bibinfo
  {journal} {Phys. Rev. B}\ }\textbf {\bibinfo {volume} {88}},\ \bibinfo
  {pages} {014206} (\bibinfo {year} {2013})}\BibitemShut {NoStop}%
\bibitem [{\citenamefont {Parameswaran}\ and\ \citenamefont
  {Vasseur}(2018)}]{Parameswaran2018}%
  \BibitemOpen
  \bibfield  {author} {\bibinfo {author} {\bibfnamefont {S.~A.}\ \bibnamefont
  {Parameswaran}}\ and\ \bibinfo {author} {\bibfnamefont {R.}~\bibnamefont
  {Vasseur}},\ }\href {https://doi.org/10.1088/1361-6633/aac9ed} {\bibfield
  {journal} {\bibinfo  {journal} {Rep. Prog. Phys.}\ }\textbf {\bibinfo
  {volume} {81}},\ \bibinfo {pages} {082501} (\bibinfo {year}
  {2018})}\BibitemShut {NoStop}%
\bibitem [{\citenamefont {Kitaev}(2001)}]{Kitaev2001}%
  \BibitemOpen
  \bibfield  {author} {\bibinfo {author} {\bibfnamefont {A.~Y.}\ \bibnamefont
  {Kitaev}},\ }\href {https://doi.org/10.1070/1063-7869/44/10S/S29} {\bibfield
  {journal} {\bibinfo  {journal} {Phys. Usp.}\ }\textbf {\bibinfo {volume}
  {44}},\ \bibinfo {pages} {131} (\bibinfo {year} {2001})}\BibitemShut
  {NoStop}%
\bibitem [{\citenamefont {Fendley}(2012)}]{Fendley_2012}%
  \BibitemOpen
  \bibfield  {author} {\bibinfo {author} {\bibfnamefont {P.}~\bibnamefont
  {Fendley}},\ }\href {https://doi.org/10.1088/1742-5468/2012/11/P11020}
  {\bibfield  {journal} {\bibinfo  {journal} {J. Stat. Mech.: Theory Exp.}\
  }\textbf {\bibinfo {volume} {2012}}\bibinfo  {number} { (11)},\ \bibinfo
  {pages} {P11020}}\BibitemShut {NoStop}%
\bibitem [{\citenamefont {Fendley}(2016)}]{Fendley_2016}%
  \BibitemOpen
\bibfield  {number} {  }\bibfield  {author} {\bibinfo {author} {\bibfnamefont
  {P.}~\bibnamefont {Fendley}},\ }\href
  {https://doi.org/10.1088/1751-8113/49/30/30LT01} {\bibfield  {journal}
  {\bibinfo  {journal} {J. Phys. A}\ }\textbf {\bibinfo {volume} {49}},\
  \bibinfo {pages} {30LT01} (\bibinfo {year} {2016})}\BibitemShut {NoStop}%
\bibitem [{\citenamefont {Kemp}\ \emph {et~al.}(2020)\citenamefont {Kemp},
  \citenamefont {Yao},\ and\ \citenamefont {Laumann}}]{Kemp2019}%
  \BibitemOpen
  \bibfield  {author} {\bibinfo {author} {\bibfnamefont {J.}~\bibnamefont
  {Kemp}}, \bibinfo {author} {\bibfnamefont {N.~Y.}\ \bibnamefont {Yao}},\ and\
  \bibinfo {author} {\bibfnamefont {C.~R.}\ \bibnamefont {Laumann}},\ }\href
  {https://doi.org/10.1103/PhysRevLett.125.200506} {\bibfield  {journal}
  {\bibinfo  {journal} {Phys. Rev. Lett.}\ }\textbf {\bibinfo {volume} {125}},\
  \bibinfo {pages} {200506} (\bibinfo {year} {2020})}\BibitemShut {NoStop}%
\bibitem [{\citenamefont {Srivatsa}\ \emph {et~al.}(2020)\citenamefont
  {Srivatsa}, \citenamefont {Wildeboer}, \citenamefont {Seidel},\ and\
  \citenamefont {Nielsen}}]{Srivatsa2020}%
  \BibitemOpen
  \bibfield  {author} {\bibinfo {author} {\bibfnamefont {N.~S.}\ \bibnamefont
  {Srivatsa}}, \bibinfo {author} {\bibfnamefont {J.}~\bibnamefont {Wildeboer}},
  \bibinfo {author} {\bibfnamefont {A.}~\bibnamefont {Seidel}},\ and\ \bibinfo
  {author} {\bibfnamefont {A.~E.~B.}\ \bibnamefont {Nielsen}},\ }\href
  {https://doi.org/10.1103/PhysRevB.102.235106} {\bibfield  {journal} {\bibinfo
   {journal} {Phys. Rev. B}\ }\textbf {\bibinfo {volume} {102}},\ \bibinfo
  {pages} {235106} (\bibinfo {year} {2020})}\BibitemShut {NoStop}%
\bibitem [{\citenamefont {Jeyaretnam}\ \emph {et~al.}(2021)\citenamefont
  {Jeyaretnam}, \citenamefont {Richter},\ and\ \citenamefont
  {Pal}}]{Jeyaretnam2021}%
  \BibitemOpen
  \bibfield  {author} {\bibinfo {author} {\bibfnamefont {J.}~\bibnamefont
  {Jeyaretnam}}, \bibinfo {author} {\bibfnamefont {J.}~\bibnamefont
  {Richter}},\ and\ \bibinfo {author} {\bibfnamefont {A.}~\bibnamefont {Pal}},\
  }\href {https://doi.org/10.1103/PhysRevB.104.014424} {\bibfield  {journal}
  {\bibinfo  {journal} {Phys. Rev. B}\ }\textbf {\bibinfo {volume} {104}},\
  \bibinfo {pages} {014424} (\bibinfo {year} {2021})}\BibitemShut {NoStop}%
\bibitem [{\citenamefont {Pekker}\ \emph {et~al.}(2014)\citenamefont {Pekker},
  \citenamefont {Refael}, \citenamefont {Altman}, \citenamefont {Demler},\ and\
  \citenamefont {Oganesyan}}]{Pekker2014}%
  \BibitemOpen
  \bibfield  {author} {\bibinfo {author} {\bibfnamefont {D.}~\bibnamefont
  {Pekker}}, \bibinfo {author} {\bibfnamefont {G.}~\bibnamefont {Refael}},
  \bibinfo {author} {\bibfnamefont {E.}~\bibnamefont {Altman}}, \bibinfo
  {author} {\bibfnamefont {E.}~\bibnamefont {Demler}},\ and\ \bibinfo {author}
  {\bibfnamefont {V.}~\bibnamefont {Oganesyan}},\ }\href
  {https://doi.org/10.1103/PhysRevX.4.011052} {\bibfield  {journal} {\bibinfo
  {journal} {Phys. Rev. X}\ }\textbf {\bibinfo {volume} {4}},\ \bibinfo {pages}
  {011052} (\bibinfo {year} {2014})}\BibitemShut {NoStop}%
\bibitem [{\citenamefont {Kj\"all}\ \emph {et~al.}(2014)\citenamefont
  {Kj\"all}, \citenamefont {Bardarson},\ and\ \citenamefont
  {Pollmann}}]{Kjall2014}%
  \BibitemOpen
  \bibfield  {author} {\bibinfo {author} {\bibfnamefont {J.~A.}\ \bibnamefont
  {Kj\"all}}, \bibinfo {author} {\bibfnamefont {J.~H.}\ \bibnamefont
  {Bardarson}},\ and\ \bibinfo {author} {\bibfnamefont {F.}~\bibnamefont
  {Pollmann}},\ }\href {https://doi.org/10.1103/PhysRevLett.113.107204}
  {\bibfield  {journal} {\bibinfo  {journal} {Phys. Rev. Lett.}\ }\textbf
  {\bibinfo {volume} {113}},\ \bibinfo {pages} {107204} (\bibinfo {year}
  {2014})}\BibitemShut {NoStop}%
\bibitem [{\citenamefont {Sahay}\ \emph {et~al.}(2021)\citenamefont {Sahay},
  \citenamefont {Machado}, \citenamefont {Ye}, \citenamefont {Laumann},\ and\
  \citenamefont {Yao}}]{Sahay2021}%
  \BibitemOpen
  \bibfield  {author} {\bibinfo {author} {\bibfnamefont {R.}~\bibnamefont
  {Sahay}}, \bibinfo {author} {\bibfnamefont {F.}~\bibnamefont {Machado}},
  \bibinfo {author} {\bibfnamefont {B.}~\bibnamefont {Ye}}, \bibinfo {author}
  {\bibfnamefont {C.~R.}\ \bibnamefont {Laumann}},\ and\ \bibinfo {author}
  {\bibfnamefont {N.~Y.}\ \bibnamefont {Yao}},\ }\href
  {https://doi.org/10.1103/PhysRevLett.126.100604} {\bibfield  {journal}
  {\bibinfo  {journal} {Phys. Rev. Lett.}\ }\textbf {\bibinfo {volume} {126}},\
  \bibinfo {pages} {100604} (\bibinfo {year} {2021})}\BibitemShut {NoStop}%
\bibitem [{\citenamefont {Moudgalya}\ \emph {et~al.}(2020)\citenamefont
  {Moudgalya}, \citenamefont {Huse},\ and\ \citenamefont
  {Khemani}}]{Moudgalya2020b}%
  \BibitemOpen
  \bibfield  {author} {\bibinfo {author} {\bibfnamefont {S.}~\bibnamefont
  {Moudgalya}}, \bibinfo {author} {\bibfnamefont {D.~A.}\ \bibnamefont
  {Huse}},\ and\ \bibinfo {author} {\bibfnamefont {V.}~\bibnamefont
  {Khemani}},\ }\Eprint {https://arxiv.org/abs/2008.09113} {arXiv:2008.09113
  [cond-mat.dis-nn]}  (\bibinfo {year} {2020})\BibitemShut {NoStop}%
\bibitem [{\citenamefont {Laflorencie}\ \emph {et~al.}(2022)\citenamefont
  {Laflorencie}, \citenamefont {Lemari\'e},\ and\ \citenamefont
  {Mac\'e}}]{Laflorencie2022}%
  \BibitemOpen
  \bibfield  {author} {\bibinfo {author} {\bibfnamefont {N.}~\bibnamefont
  {Laflorencie}}, \bibinfo {author} {\bibfnamefont {G.}~\bibnamefont
  {Lemari\'e}},\ and\ \bibinfo {author} {\bibfnamefont {N.}~\bibnamefont
  {Mac\'e}},\ }\href {https://doi.org/10.1103/PhysRevResearch.4.L032016}
  {\bibfield  {journal} {\bibinfo  {journal} {Phys. Rev. Res.}\ }\textbf
  {\bibinfo {volume} {4}},\ \bibinfo {pages} {L032016} (\bibinfo {year}
  {2022})}\BibitemShut {NoStop}%
\bibitem [{\citenamefont {Agarwal}\ \emph {et~al.}(2017)\citenamefont
  {Agarwal}, \citenamefont {Altman}, \citenamefont {Demler}, \citenamefont
  {Gopalakrishnan}, \citenamefont {Huse},\ and\ \citenamefont
  {Knap}}]{Agarwal2017}%
  \BibitemOpen
  \bibfield  {author} {\bibinfo {author} {\bibfnamefont {K.}~\bibnamefont
  {Agarwal}}, \bibinfo {author} {\bibfnamefont {E.}~\bibnamefont {Altman}},
  \bibinfo {author} {\bibfnamefont {E.}~\bibnamefont {Demler}}, \bibinfo
  {author} {\bibfnamefont {S.}~\bibnamefont {Gopalakrishnan}}, \bibinfo
  {author} {\bibfnamefont {D.~A.}\ \bibnamefont {Huse}},\ and\ \bibinfo
  {author} {\bibfnamefont {M.}~\bibnamefont {Knap}},\ }\href
  {https://doi.org/https://doi.org/10.1002/andp.201600326} {\bibfield
  {journal} {\bibinfo  {journal} {Ann. Phys. (Leipzig)}\ }\textbf {\bibinfo
  {volume} {529}},\ \bibinfo {pages} {1600326} (\bibinfo {year}
  {2017})}\BibitemShut {NoStop}%
\bibitem [{\citenamefont {Luitz}\ \emph {et~al.}(2017)\citenamefont {Luitz},
  \citenamefont {Huveneers},\ and\ \citenamefont {De~Roeck}}]{Luitz2017}%
  \BibitemOpen
  \bibfield  {author} {\bibinfo {author} {\bibfnamefont {D.~J.}\ \bibnamefont
  {Luitz}}, \bibinfo {author} {\bibfnamefont {F.}~\bibnamefont {Huveneers}},\
  and\ \bibinfo {author} {\bibfnamefont {W.}~\bibnamefont {De~Roeck}},\ }\href
  {https://doi.org/10.1103/PhysRevLett.119.150602} {\bibfield  {journal}
  {\bibinfo  {journal} {Phys. Rev. Lett.}\ }\textbf {\bibinfo {volume} {119}},\
  \bibinfo {pages} {150602} (\bibinfo {year} {2017})}\BibitemShut {NoStop}%
\bibitem [{\citenamefont {De~Roeck}\ and\ \citenamefont
  {Huveneers}(2017)}]{DeRoeck2017}%
  \BibitemOpen
  \bibfield  {author} {\bibinfo {author} {\bibfnamefont {W.}~\bibnamefont
  {De~Roeck}}\ and\ \bibinfo {author} {\bibfnamefont {F.}~\bibnamefont
  {Huveneers}},\ }\href {https://doi.org/10.1103/PhysRevB.95.155129} {\bibfield
   {journal} {\bibinfo  {journal} {Phys. Rev. B}\ }\textbf {\bibinfo {volume}
  {95}},\ \bibinfo {pages} {155129} (\bibinfo {year} {2017})}\BibitemShut
  {NoStop}%
\bibitem [{\citenamefont {\v{S}untajs}\ \emph {et~al.}(2020)\citenamefont
  {\v{S}untajs}, \citenamefont {Bon\v{c}a}, \citenamefont {Prosen},\ and\
  \citenamefont {Vidmar}}]{Suntajs2020}%
  \BibitemOpen
  \bibfield  {author} {\bibinfo {author} {\bibfnamefont {J.}~\bibnamefont
  {\v{S}untajs}}, \bibinfo {author} {\bibfnamefont {J.}~\bibnamefont
  {Bon\v{c}a}}, \bibinfo {author} {\bibfnamefont {T.}~\bibnamefont {Prosen}},\
  and\ \bibinfo {author} {\bibfnamefont {L.}~\bibnamefont {Vidmar}},\ }\href
  {https://doi.org/10.1103/PhysRevE.102.062144} {\bibfield  {journal} {\bibinfo
   {journal} {Phys. Rev. E}\ }\textbf {\bibinfo {volume} {102}},\ \bibinfo
  {pages} {062144} (\bibinfo {year} {2020})}\BibitemShut {NoStop}%
\bibitem [{\citenamefont {Crowley}\ and\ \citenamefont
  {Chandran}(2022)}]{Crowley2022b}%
  \BibitemOpen
  \bibfield  {author} {\bibinfo {author} {\bibfnamefont {P.~J.~D.}\
  \bibnamefont {Crowley}}\ and\ \bibinfo {author} {\bibfnamefont
  {A.}~\bibnamefont {Chandran}},\ }\href
  {https://doi.org/10.1103/PhysRevB.106.184208} {\bibfield  {journal} {\bibinfo
   {journal} {Phys. Rev. B}\ }\textbf {\bibinfo {volume} {106}},\ \bibinfo
  {pages} {184208} (\bibinfo {year} {2022})}\BibitemShut {NoStop}%
\bibitem [{\citenamefont {Sels}(2022)}]{Sels2022}%
  \BibitemOpen
  \bibfield  {author} {\bibinfo {author} {\bibfnamefont {D.}~\bibnamefont
  {Sels}},\ }\href {https://doi.org/10.1103/PhysRevB.106.L020202} {\bibfield
  {journal} {\bibinfo  {journal} {Phys. Rev. B}\ }\textbf {\bibinfo {volume}
  {106}},\ \bibinfo {pages} {L020202} (\bibinfo {year} {2022})}\BibitemShut
  {NoStop}%
\bibitem [{\citenamefont {Tikhonov}\ and\ \citenamefont
  {Mirlin}(2021)}]{Tikhonov2021}%
  \BibitemOpen
  \bibfield  {author} {\bibinfo {author} {\bibfnamefont {K.~S.}\ \bibnamefont
  {Tikhonov}}\ and\ \bibinfo {author} {\bibfnamefont {A.~D.}\ \bibnamefont
  {Mirlin}},\ }\href {https://doi.org/10.1016/j.aop.2021.168525} {\bibfield
  {journal} {\bibinfo  {journal} {Ann. Phys. (NY)}\ }\textbf {\bibinfo {volume}
  {435}},\ \bibinfo {pages} {168525} (\bibinfo {year} {2021})}\BibitemShut
  {NoStop}%
\bibitem [{\citenamefont {Morningstar}\ \emph {et~al.}(2022)\citenamefont
  {Morningstar}, \citenamefont {Colmenarez}, \citenamefont {Khemani},
  \citenamefont {Luitz},\ and\ \citenamefont {Huse}}]{Morningstar2022}%
  \BibitemOpen
  \bibfield  {author} {\bibinfo {author} {\bibfnamefont {A.}~\bibnamefont
  {Morningstar}}, \bibinfo {author} {\bibfnamefont {L.}~\bibnamefont
  {Colmenarez}}, \bibinfo {author} {\bibfnamefont {V.}~\bibnamefont {Khemani}},
  \bibinfo {author} {\bibfnamefont {D.~J.}\ \bibnamefont {Luitz}},\ and\
  \bibinfo {author} {\bibfnamefont {D.~A.}\ \bibnamefont {Huse}},\ }\href
  {https://doi.org/10.1103/PhysRevB.105.174205} {\bibfield  {journal} {\bibinfo
   {journal} {Phys. Rev. B}\ }\textbf {\bibinfo {volume} {105}},\ \bibinfo
  {pages} {174205} (\bibinfo {year} {2022})}\BibitemShut {NoStop}%
\bibitem [{\citenamefont {Long}\ \emph {et~al.}(2022)\citenamefont {Long},
  \citenamefont {Crowley}, \citenamefont {Khemani},\ and\ \citenamefont
  {Chandran}}]{Long2022}%
  \BibitemOpen
  \bibfield  {author} {\bibinfo {author} {\bibfnamefont {D.~M.}\ \bibnamefont
  {Long}}, \bibinfo {author} {\bibfnamefont {P.~J.~D.}\ \bibnamefont
  {Crowley}}, \bibinfo {author} {\bibfnamefont {V.}~\bibnamefont {Khemani}},\
  and\ \bibinfo {author} {\bibfnamefont {A.}~\bibnamefont {Chandran}},\
  }\Eprint {https://arxiv.org/abs/2207.05761} {arXiv:2207.05761
  [cond-mat.dis-nn]}  (\bibinfo {year} {2022})\BibitemShut {NoStop}%
\bibitem [{\citenamefont {Ha}\ \emph {et~al.}(2023)\citenamefont {Ha},
  \citenamefont {Morningstar},\ and\ \citenamefont {Huse}}]{Ha2023}%
  \BibitemOpen
  \bibfield  {author} {\bibinfo {author} {\bibfnamefont {H.}~\bibnamefont
  {Ha}}, \bibinfo {author} {\bibfnamefont {A.}~\bibnamefont {Morningstar}},\
  and\ \bibinfo {author} {\bibfnamefont {D.~A.}\ \bibnamefont {Huse}},\ }\href
  {https://doi.org/10.1103/PhysRevLett.130.250405} {\bibfield  {journal}
  {\bibinfo  {journal} {Phys. Rev. Lett.}\ }\textbf {\bibinfo {volume} {130}},\
  \bibinfo {pages} {250405} (\bibinfo {year} {2023})}\BibitemShut {NoStop}%
\bibitem [{\citenamefont {Pollmann}\ \emph {et~al.}(2016)\citenamefont
  {Pollmann}, \citenamefont {Khemani}, \citenamefont {Cirac},\ and\
  \citenamefont {Sondhi}}]{Pollman2016}%
  \BibitemOpen
  \bibfield  {author} {\bibinfo {author} {\bibfnamefont {F.}~\bibnamefont
  {Pollmann}}, \bibinfo {author} {\bibfnamefont {V.}~\bibnamefont {Khemani}},
  \bibinfo {author} {\bibfnamefont {J.~I.}\ \bibnamefont {Cirac}},\ and\
  \bibinfo {author} {\bibfnamefont {S.~L.}\ \bibnamefont {Sondhi}},\ }\href
  {https://doi.org/10.1103/PhysRevB.94.041116} {\bibfield  {journal} {\bibinfo
  {journal} {Phys. Rev. B}\ }\textbf {\bibinfo {volume} {94}},\ \bibinfo
  {pages} {041116(R)} (\bibinfo {year} {2016})}\BibitemShut {NoStop}%
\bibitem [{\citenamefont {Wahl}\ \emph {et~al.}(2017)\citenamefont {Wahl},
  \citenamefont {Pal},\ and\ \citenamefont {Simon}}]{Wahl2017}%
  \BibitemOpen
  \bibfield  {author} {\bibinfo {author} {\bibfnamefont {T.~B.}\ \bibnamefont
  {Wahl}}, \bibinfo {author} {\bibfnamefont {A.}~\bibnamefont {Pal}},\ and\
  \bibinfo {author} {\bibfnamefont {S.~H.}\ \bibnamefont {Simon}},\ }\href
  {https://doi.org/10.1103/PhysRevX.7.021018} {\bibfield  {journal} {\bibinfo
  {journal} {Phys. Rev. X}\ }\textbf {\bibinfo {volume} {7}},\ \bibinfo {pages}
  {021018} (\bibinfo {year} {2017})}\BibitemShut {NoStop}%
\bibitem [{\citenamefont {Pekker}\ and\ \citenamefont
  {Clark}(2017)}]{Pekker2017b}%
  \BibitemOpen
  \bibfield  {author} {\bibinfo {author} {\bibfnamefont {D.}~\bibnamefont
  {Pekker}}\ and\ \bibinfo {author} {\bibfnamefont {B.~K.}\ \bibnamefont
  {Clark}},\ }\href {https://doi.org/10.1103/PhysRevB.95.035116} {\bibfield
  {journal} {\bibinfo  {journal} {Phys. Rev. B}\ }\textbf {\bibinfo {volume}
  {95}},\ \bibinfo {pages} {035116} (\bibinfo {year} {2017})}\BibitemShut
  {NoStop}%
\bibitem [{\citenamefont {Wahl}\ \emph {et~al.}(2022)\citenamefont {Wahl},
  \citenamefont {Venn},\ and\ \citenamefont {B\'eri}}]{Wahl2022}%
  \BibitemOpen
  \bibfield  {author} {\bibinfo {author} {\bibfnamefont {T.~B.}\ \bibnamefont
  {Wahl}}, \bibinfo {author} {\bibfnamefont {F.}~\bibnamefont {Venn}},\ and\
  \bibinfo {author} {\bibfnamefont {B.}~\bibnamefont {B\'eri}},\ }\href
  {https://doi.org/10.1103/PhysRevB.105.144205} {\bibfield  {journal} {\bibinfo
   {journal} {Phys. Rev. B}\ }\textbf {\bibinfo {volume} {105}},\ \bibinfo
  {pages} {144205} (\bibinfo {year} {2022})}\BibitemShut {NoStop}%
\bibitem [{\citenamefont {Pekker}\ \emph {et~al.}(2017)\citenamefont {Pekker},
  \citenamefont {Clark}, \citenamefont {Oganesyan},\ and\ \citenamefont
  {Refael}}]{Pekker2017}%
  \BibitemOpen
  \bibfield  {author} {\bibinfo {author} {\bibfnamefont {D.}~\bibnamefont
  {Pekker}}, \bibinfo {author} {\bibfnamefont {B.~K.}\ \bibnamefont {Clark}},
  \bibinfo {author} {\bibfnamefont {V.}~\bibnamefont {Oganesyan}},\ and\
  \bibinfo {author} {\bibfnamefont {G.}~\bibnamefont {Refael}},\ }\href
  {https://doi.org/10.1103/PhysRevLett.119.075701} {\bibfield  {journal}
  {\bibinfo  {journal} {Phys. Rev. Lett.}\ }\textbf {\bibinfo {volume} {119}},\
  \bibinfo {pages} {075701} (\bibinfo {year} {2017})}\BibitemShut {NoStop}%
\bibitem [{\citenamefont {Vosk}\ and\ \citenamefont {Altman}(2013)}]{Vosk2013}%
  \BibitemOpen
  \bibfield  {author} {\bibinfo {author} {\bibfnamefont {R.}~\bibnamefont
  {Vosk}}\ and\ \bibinfo {author} {\bibfnamefont {E.}~\bibnamefont {Altman}},\
  }\href {https://doi.org/10.1103/PhysRevLett.110.067204} {\bibfield  {journal}
  {\bibinfo  {journal} {Phys. Rev. Lett.}\ }\textbf {\bibinfo {volume} {110}},\
  \bibinfo {pages} {067204} (\bibinfo {year} {2013})}\BibitemShut {NoStop}%
\bibitem [{\citenamefont {Vasseur}\ \emph {et~al.}(2015)\citenamefont
  {Vasseur}, \citenamefont {Potter},\ and\ \citenamefont
  {Parameswaran}}]{Vasseur2015}%
  \BibitemOpen
  \bibfield  {author} {\bibinfo {author} {\bibfnamefont {R.}~\bibnamefont
  {Vasseur}}, \bibinfo {author} {\bibfnamefont {A.~C.}\ \bibnamefont
  {Potter}},\ and\ \bibinfo {author} {\bibfnamefont {S.~A.}\ \bibnamefont
  {Parameswaran}},\ }\href {https://doi.org/10.1103/PhysRevLett.114.217201}
  {\bibfield  {journal} {\bibinfo  {journal} {Phys. Rev. Lett.}\ }\textbf
  {\bibinfo {volume} {114}},\ \bibinfo {pages} {217201} (\bibinfo {year}
  {2015})}\BibitemShut {NoStop}%
\bibitem [{\citenamefont {Goremykina}\ \emph {et~al.}(2019)\citenamefont
  {Goremykina}, \citenamefont {Vasseur},\ and\ \citenamefont
  {Serbyn}}]{Goremykina2019}%
  \BibitemOpen
  \bibfield  {author} {\bibinfo {author} {\bibfnamefont {A.}~\bibnamefont
  {Goremykina}}, \bibinfo {author} {\bibfnamefont {R.}~\bibnamefont
  {Vasseur}},\ and\ \bibinfo {author} {\bibfnamefont {M.}~\bibnamefont
  {Serbyn}},\ }\href {https://doi.org/10.1103/PhysRevLett.122.040601}
  {\bibfield  {journal} {\bibinfo  {journal} {Phys. Rev. Lett.}\ }\textbf
  {\bibinfo {volume} {122}},\ \bibinfo {pages} {040601} (\bibinfo {year}
  {2019})}\BibitemShut {NoStop}%
\bibitem [{\citenamefont {Dasgupta}\ and\ \citenamefont
  {Ma}(1980)}]{Dasgupta1980}%
  \BibitemOpen
  \bibfield  {author} {\bibinfo {author} {\bibfnamefont {C.}~\bibnamefont
  {Dasgupta}}\ and\ \bibinfo {author} {\bibfnamefont {S.~K.}\ \bibnamefont
  {Ma}},\ }\href {https://doi.org/10.1103/PhysRevB.22.1305} {\bibfield
  {journal} {\bibinfo  {journal} {Phys. Rev. B}\ }\textbf {\bibinfo {volume}
  {22}},\ \bibinfo {pages} {1305} (\bibinfo {year} {1980})}\BibitemShut
  {NoStop}%
\bibitem [{\citenamefont {Fisher}(1992)}]{Fisher1992}%
  \BibitemOpen
  \bibfield  {author} {\bibinfo {author} {\bibfnamefont {D.~S.}\ \bibnamefont
  {Fisher}},\ }\href {https://doi.org/10.1103/PhysRevLett.69.534} {\bibfield
  {journal} {\bibinfo  {journal} {Phys. Rev. Lett.}\ }\textbf {\bibinfo
  {volume} {69}},\ \bibinfo {pages} {534} (\bibinfo {year} {1992})}\BibitemShut
  {NoStop}%
\bibitem [{\citenamefont {You}\ \emph {et~al.}(2016)\citenamefont {You},
  \citenamefont {Qi},\ and\ \citenamefont {Xu}}]{You2016}%
  \BibitemOpen
  \bibfield  {author} {\bibinfo {author} {\bibfnamefont {Y.-Z.}\ \bibnamefont
  {You}}, \bibinfo {author} {\bibfnamefont {X.-L.}\ \bibnamefont {Qi}},\ and\
  \bibinfo {author} {\bibfnamefont {C.}~\bibnamefont {Xu}},\ }\href
  {https://doi.org/10.1103/PhysRevB.93.104205} {\bibfield  {journal} {\bibinfo
  {journal} {Phys. Rev. B}\ }\textbf {\bibinfo {volume} {93}},\ \bibinfo
  {pages} {104205} (\bibinfo {year} {2016})}\BibitemShut {NoStop}%
\bibitem [{\citenamefont {Potter}\ \emph {et~al.}(2015)\citenamefont {Potter},
  \citenamefont {Vasseur},\ and\ \citenamefont {Parameswaran}}]{Potter2015}%
  \BibitemOpen
  \bibfield  {author} {\bibinfo {author} {\bibfnamefont {A.~C.}\ \bibnamefont
  {Potter}}, \bibinfo {author} {\bibfnamefont {R.}~\bibnamefont {Vasseur}},\
  and\ \bibinfo {author} {\bibfnamefont {S.~A.}\ \bibnamefont {Parameswaran}},\
  }\href {https://doi.org/10.1103/PhysRevX.5.031033} {\bibfield  {journal}
  {\bibinfo  {journal} {Phys. Rev. X}\ }\textbf {\bibinfo {volume} {5}},\
  \bibinfo {pages} {031033} (\bibinfo {year} {2015})}\BibitemShut {NoStop}%
\bibitem [{\citenamefont {Roeck}\ and\ \citenamefont
  {Imbrie}(2017)}]{DeRoeck2017b}%
  \BibitemOpen
  \bibfield  {author} {\bibinfo {author} {\bibfnamefont {W.~D.}\ \bibnamefont
  {Roeck}}\ and\ \bibinfo {author} {\bibfnamefont {J.~Z.}\ \bibnamefont
  {Imbrie}},\ }\href {https://doi.org/10.1098/rsta.2016.0422} {\bibfield
  {journal} {\bibinfo  {journal} {Phil. Trans. R. Soc. A}\ }\textbf {\bibinfo
  {volume} {375}},\ \bibinfo {pages} {20160422} (\bibinfo {year}
  {2017})}\BibitemShut {NoStop}%
\bibitem [{\citenamefont {Protopopov}\ \emph {et~al.}(2017)\citenamefont
  {Protopopov}, \citenamefont {Ho},\ and\ \citenamefont
  {Abanin}}]{Protopopov2017}%
  \BibitemOpen
  \bibfield  {author} {\bibinfo {author} {\bibfnamefont {I.~V.}\ \bibnamefont
  {Protopopov}}, \bibinfo {author} {\bibfnamefont {W.~W.}\ \bibnamefont {Ho}},\
  and\ \bibinfo {author} {\bibfnamefont {D.~A.}\ \bibnamefont {Abanin}},\
  }\href {https://doi.org/10.1103/PhysRevB.96.041122} {\bibfield  {journal}
  {\bibinfo  {journal} {Phys. Rev. B}\ }\textbf {\bibinfo {volume} {96}},\
  \bibinfo {pages} {041122(R)} (\bibinfo {year} {2017})}\BibitemShut {NoStop}%
\bibitem [{\citenamefont {Gornyi}\ \emph {et~al.}(2017)\citenamefont {Gornyi},
  \citenamefont {Mirlin}, \citenamefont {Polyakov},\ and\ \citenamefont
  {Burin}}]{Gornyi2017}%
  \BibitemOpen
  \bibfield  {author} {\bibinfo {author} {\bibfnamefont {I.~V.}\ \bibnamefont
  {Gornyi}}, \bibinfo {author} {\bibfnamefont {A.~D.}\ \bibnamefont {Mirlin}},
  \bibinfo {author} {\bibfnamefont {D.~G.}\ \bibnamefont {Polyakov}},\ and\
  \bibinfo {author} {\bibfnamefont {A.~L.}\ \bibnamefont {Burin}},\ }\href
  {https://doi.org/10.1002/andp.201600360} {\bibfield  {journal} {\bibinfo
  {journal} {Ann. Phys. (Leipzig)}\ }\textbf {\bibinfo {volume} {529}},\
  \bibinfo {pages} {1600360} (\bibinfo {year} {2017})}\BibitemShut {NoStop}%
\bibitem [{\citenamefont {Protopopov}\ \emph {et~al.}(2020)\citenamefont
  {Protopopov}, \citenamefont {Panda}, \citenamefont {Parolini}, \citenamefont
  {Scardicchio}, \citenamefont {Demler},\ and\ \citenamefont
  {Abanin}}]{Protopopov2020}%
  \BibitemOpen
  \bibfield  {author} {\bibinfo {author} {\bibfnamefont {I.~V.}\ \bibnamefont
  {Protopopov}}, \bibinfo {author} {\bibfnamefont {R.~K.}\ \bibnamefont
  {Panda}}, \bibinfo {author} {\bibfnamefont {T.}~\bibnamefont {Parolini}},
  \bibinfo {author} {\bibfnamefont {A.}~\bibnamefont {Scardicchio}}, \bibinfo
  {author} {\bibfnamefont {E.}~\bibnamefont {Demler}},\ and\ \bibinfo {author}
  {\bibfnamefont {D.~A.}\ \bibnamefont {Abanin}},\ }\href
  {https://doi.org/10.1103/PhysRevX.10.011025} {\bibfield  {journal} {\bibinfo
  {journal} {Phys. Rev. X}\ }\textbf {\bibinfo {volume} {10}},\ \bibinfo
  {pages} {011025} (\bibinfo {year} {2020})}\BibitemShut {NoStop}%
\bibitem [{\citenamefont {Kemp}\ \emph {et~al.}(2017)\citenamefont {Kemp},
  \citenamefont {Yao}, \citenamefont {Laumann},\ and\ \citenamefont
  {Fendley}}]{Kemp_2017}%
  \BibitemOpen
  \bibfield  {author} {\bibinfo {author} {\bibfnamefont {J.}~\bibnamefont
  {Kemp}}, \bibinfo {author} {\bibfnamefont {N.~Y.}\ \bibnamefont {Yao}},
  \bibinfo {author} {\bibfnamefont {C.~R.}\ \bibnamefont {Laumann}},\ and\
  \bibinfo {author} {\bibfnamefont {P.}~\bibnamefont {Fendley}},\ }\href
  {https://doi.org/10.1088/1742-5468/aa73f0} {\bibfield  {journal} {\bibinfo
  {journal} {J. Stat. Mech.: Theory Exp.}\ }\textbf {\bibinfo {volume}
  {2017}}\bibinfo  {number} { (6)},\ \bibinfo {pages} {063105}}\BibitemShut
  {NoStop}%
\bibitem [{\citenamefont {Karcher}\ \emph {et~al.}(2019)\citenamefont
  {Karcher}, \citenamefont {Sonner},\ and\ \citenamefont
  {Mirlin}}]{Karcher2019}%
  \BibitemOpen
\bibfield  {number} {  }\bibfield  {author} {\bibinfo {author} {\bibfnamefont
  {J.~F.}\ \bibnamefont {Karcher}}, \bibinfo {author} {\bibfnamefont
  {M.}~\bibnamefont {Sonner}},\ and\ \bibinfo {author} {\bibfnamefont {A.~D.}\
  \bibnamefont {Mirlin}},\ }\href {https://doi.org/10.1103/PhysRevB.100.134207}
  {\bibfield  {journal} {\bibinfo  {journal} {Phys. Rev. B}\ }\textbf {\bibinfo
  {volume} {100}},\ \bibinfo {pages} {134207} (\bibinfo {year}
  {2019})}\BibitemShut {NoStop}%
\bibitem [{\citenamefont {Schrieffer}\ and\ \citenamefont
  {Wolff}(1966)}]{Schrieffer1966}%
  \BibitemOpen
  \bibfield  {author} {\bibinfo {author} {\bibfnamefont {J.~R.}\ \bibnamefont
  {Schrieffer}}\ and\ \bibinfo {author} {\bibfnamefont {P.~A.}\ \bibnamefont
  {Wolff}},\ }\href {https://doi.org/10.1103/PhysRev.149.491} {\bibfield
  {journal} {\bibinfo  {journal} {Phys. Rev.}\ }\textbf {\bibinfo {volume}
  {149}},\ \bibinfo {pages} {491} (\bibinfo {year} {1966})}\BibitemShut
  {NoStop}%
\bibitem [{\citenamefont {Ware}\ \emph {et~al.}(2021)\citenamefont {Ware},
  \citenamefont {Abanin},\ and\ \citenamefont {Vasseur}}]{Ware2021}%
  \BibitemOpen
  \bibfield  {author} {\bibinfo {author} {\bibfnamefont {B.}~\bibnamefont
  {Ware}}, \bibinfo {author} {\bibfnamefont {D.}~\bibnamefont {Abanin}},\ and\
  \bibinfo {author} {\bibfnamefont {R.}~\bibnamefont {Vasseur}},\ }\href
  {https://doi.org/10.1103/PhysRevB.103.094203} {\bibfield  {journal} {\bibinfo
   {journal} {Phys. Rev. B}\ }\textbf {\bibinfo {volume} {103}},\ \bibinfo
  {pages} {094203} (\bibinfo {year} {2021})}\BibitemShut {NoStop}%
\bibitem [{\citenamefont {Dehaene}\ and\ \citenamefont
  {DeMoor}(2003)}]{Dehaene2003}%
  \BibitemOpen
  \bibfield  {author} {\bibinfo {author} {\bibfnamefont {J.}~\bibnamefont
  {Dehaene}}\ and\ \bibinfo {author} {\bibfnamefont {B.}~\bibnamefont
  {DeMoor}},\ }\href {https://doi.org/10.1103/PhysRevA.68.042318} {\bibfield
  {journal} {\bibinfo  {journal} {Phys. Rev. A}\ }\textbf {\bibinfo {volume}
  {68}},\ \bibinfo {pages} {042318} (\bibinfo {year} {2003})}\BibitemShut
  {NoStop}%
\bibitem [{\citenamefont {Aaronson}\ and\ \citenamefont
  {Gottesman}(2004)}]{Aaronson2004}%
  \BibitemOpen
  \bibfield  {author} {\bibinfo {author} {\bibfnamefont {S.}~\bibnamefont
  {Aaronson}}\ and\ \bibinfo {author} {\bibfnamefont {D.}~\bibnamefont
  {Gottesman}},\ }\href {https://doi.org/10.1103/PhysRevA.70.052328} {\bibfield
   {journal} {\bibinfo  {journal} {Phys. Rev. A}\ }\textbf {\bibinfo {volume}
  {70}},\ \bibinfo {pages} {052328} (\bibinfo {year} {2004})}\BibitemShut
  {NoStop}%
\bibitem [{\citenamefont {Gottesman}(1997)}]{Gottesman1997}%
  \BibitemOpen
  \bibfield  {author} {\bibinfo {author} {\bibfnamefont {D.}~\bibnamefont
  {Gottesman}},\ }\emph {\bibinfo {title} {Stabilizer codes and quantum error
  correction}},\ \href@noop {} {Ph.D. thesis},\ \bibinfo  {school} {California
  Institute of Technology} (\bibinfo {year} {1997})\BibitemShut {NoStop}%
\bibitem [{\citenamefont {Vidal}(2008)}]{Vidal2008}%
  \BibitemOpen
  \bibfield  {author} {\bibinfo {author} {\bibfnamefont {G.}~\bibnamefont
  {Vidal}},\ }\href {https://doi.org/10.1103/PhysRevLett.101.110501} {\bibfield
   {journal} {\bibinfo  {journal} {Phys. Rev. Lett.}\ }\textbf {\bibinfo
  {volume} {101}},\ \bibinfo {pages} {110501} (\bibinfo {year}
  {2008})}\BibitemShut {NoStop}%
\bibitem [{\citenamefont {Serbyn}\ \emph {et~al.}(2015)\citenamefont {Serbyn},
  \citenamefont {Papi\'c},\ and\ \citenamefont {Abanin}}]{Serbyn2015}%
  \BibitemOpen
  \bibfield  {author} {\bibinfo {author} {\bibfnamefont {M.}~\bibnamefont
  {Serbyn}}, \bibinfo {author} {\bibfnamefont {Z.}~\bibnamefont {Papi\'c}},\
  and\ \bibinfo {author} {\bibfnamefont {D.~A.}\ \bibnamefont {Abanin}},\
  }\href {https://doi.org/10.1103/PhysRevX.5.041047} {\bibfield  {journal}
  {\bibinfo  {journal} {Phys. Rev. X}\ }\textbf {\bibinfo {volume} {5}},\
  \bibinfo {pages} {041047} (\bibinfo {year} {2015})}\BibitemShut {NoStop}%
\bibitem [{\citenamefont {Kiefer-Emmanouilidis}\ \emph
  {et~al.}(2021)\citenamefont {Kiefer-Emmanouilidis}, \citenamefont {Unanyan},
  \citenamefont {Fleischhauer},\ and\ \citenamefont
  {Sirker}}]{Emmanouilidis2021}%
  \BibitemOpen
  \bibfield  {author} {\bibinfo {author} {\bibfnamefont {M.}~\bibnamefont
  {Kiefer-Emmanouilidis}}, \bibinfo {author} {\bibfnamefont {R.}~\bibnamefont
  {Unanyan}}, \bibinfo {author} {\bibfnamefont {M.}~\bibnamefont
  {Fleischhauer}},\ and\ \bibinfo {author} {\bibfnamefont {J.}~\bibnamefont
  {Sirker}},\ }\href {https://doi.org/10.1103/PhysRevB.103.024203} {\bibfield
  {journal} {\bibinfo  {journal} {Phys. Rev. B}\ }\textbf {\bibinfo {volume}
  {103}},\ \bibinfo {pages} {024203} (\bibinfo {year} {2021})}\BibitemShut
  {NoStop}%
\bibitem [{\citenamefont {Varvelis}\ and\ \citenamefont
  {DiVincenzo}(2022)}]{Varvelis2022}%
  \BibitemOpen
  \bibfield  {author} {\bibinfo {author} {\bibfnamefont {E.}~\bibnamefont
  {Varvelis}}\ and\ \bibinfo {author} {\bibfnamefont {D.~P.}\ \bibnamefont
  {DiVincenzo}},\ }\Eprint {https://arxiv.org/abs/2212.03805} {arXiv:2212.03805
  [quant-ph]}  (\bibinfo {year} {2022})\BibitemShut {NoStop}%
\bibitem [{\citenamefont {Gottlob}\ and\ \citenamefont
  {Schneider}(2023)}]{Gottlob2022}%
  \BibitemOpen
  \bibfield  {author} {\bibinfo {author} {\bibfnamefont {E.}~\bibnamefont
  {Gottlob}}\ and\ \bibinfo {author} {\bibfnamefont {U.}~\bibnamefont
  {Schneider}},\ }\href {https://doi.org/10.1103/PhysRevB.107.144202}
  {\bibfield  {journal} {\bibinfo  {journal} {Phys. Rev. B}\ }\textbf {\bibinfo
  {volume} {107}},\ \bibinfo {pages} {144202} (\bibinfo {year}
  {2023})}\BibitemShut {NoStop}%
\end{thebibliography}%
\end{document}